\documentclass[a4paper,aps,twocolumn,showpacs,showkeys,nofootinbib,preprintnumbers,superscriptaddress,amsmath,amssymb,amsfonts]{revtex4-1}
\usepackage{graphicx}
\usepackage{natbib,times,multirow,color}
\usepackage{hyperref}
\hypersetup{
  colorlinks=true,        
  linkcolor=blue,         
  citecolor=magenta,      
}

\newcommand{\ie}{{i.e.,}~}
\newcommand{\eg}{{e.g.,}~}

\begin{document}

\title{Test-particle dynamics in general spherically symmetric black hole spacetimes}

\author{Mariafelicia De Laurentis}
\email{laurentis@th.physik.uni-frankfurt.de}
\affiliation{Institut f\"ur Theoretische Physik, Max-von-Laue-Stra{\ss}e 1, 
D-60438 Frankfurt am Main, Germany}
 
\author{Ziri Younsi}
\email{younsi@th.physik.uni-frankfurt.de}
\affiliation{Institut f\"ur Theoretische Physik, Max-von-Laue-Stra{\ss}e 1, 
D-60438 Frankfurt am Main, Germany}

\author{Oliver Porth}
\email{porth@th.physik.uni-frankfurt.de}
\affiliation{Institut f\"ur Theoretische Physik, Max-von-Laue-Stra{\ss}e 1, 
D-60438 Frankfurt am Main, Germany}

\author{Yosuke Mizuno}
\email{mizuno@th.physik.uni-frankfurt.de}
\affiliation{Institut f\"ur Theoretische Physik, Max-von-Laue-Stra{\ss}e 1, 
D-60438 Frankfurt am Main, Germany}

\author{Luciano Rezzolla}
\email{rezzolla@th.physik.uni-frankfurt.de}	
\affiliation{Institut f\"ur Theoretische Physik, Max-von-Laue-Stra{\ss}e 1, 
D-60438 Frankfurt am Main, Germany}
\affiliation{Frankfurt Institute for Advanced Studies, Ruth-Moufang-Strasse 1, 
60438 Frankfurt, Germany}
 
\date{\today}
\begin{abstract}
To date, the most precise tests of general relativity
have been achieved through pulsar timing, albeit in the weak-field regime.
Since pulsars are some of the most precise and stable ``clocks'' in the Universe,
present observational efforts are focused on detecting pulsars in the vicinity of
supermassive black holes (most notably in our Galactic Centre), enabling pulsar timing
to be used as an extremely precise probe of strong-field gravity.
In this paper a mathematical framework to describe
test-particle dynamics in general black hole spacetimes is presented, and subsequently
used to study a binary system comprising a pulsar orbiting a black hole.
In particular, taking into account the parameterization of
a general spherically symmetric black hole metric,
general analytic expressions for both the advance of the periastron and for
the orbital period of a massive test particle are derived.
Furthermore, these expressions are applied to four
representative cases of solutions arising in both general relativity and in
alternative theories of gravity. 
Finally, this framework is applied to
the Galactic Centre $S$-stars and four distinct pulsar toy models.
It is shown that by adopting a fully general-relativistic description of
test-particle motion which is independent of any particular theory of gravity,
observations of pulsars can help impose better constraints on
alternative theories of gravity than is presently possible.
\end{abstract}



\maketitle

\section{introduction}
\label{sec:uno}

It is now widely believed that
supermassive black holes (SMBHs) reside at the centres of all galaxies and
that their estimated masses are in the range of a few million to up to
tens of billions of solar masses. 
Earth's closest SMBH candidate is found at the Galactic Centre,
Sagittarius A* (Sgr~A*), which astronomers have been observing for
several decades \cite{Eckart1996,Ghez:2008}.

It is expected that the mathematical description of astrophysical black
holes (BHs) is based on solutions to the Einstein field equations, and therefore
founded on general relativity (GR).
However, there also exist many BH solutions in extended and alternative
theories of gravity, and to-date observational constraints, most notably in the 
strong-field regime, are lacking.
Moreover, modifications or extensions of classical GR, let alone entirely
new theories of gravity, are not without astrophysical motivation 
(e.g., accounting for inflation, dark matter and dark energy) \cite[see][for a review]{Capozziello2011}.

One promising probe of strong-field gravity is the direct imaging of
the shadow cast by a SMBH.
High-resolution imaging of the event horizon and BH shadow can
improve our understanding of gravity in the strong-field regime
and, hopefully, provide direct evidence as to whether BHs exist and which
theory (or classes of theory) of gravity describe them best 
\cite{Cunningham1973, Falcke2000,
Grenzebach14, Abdujabbarov2015, Younsi2016}. 
Direct observation of the event horizon of our own
SMBH, Sgr~A*, will soon be obtained by the Event Horizon Telescope
Collaboration (EHTC)\footnote{See
\href{http://www.eventhorizontelescope.org}{www.eventhorizontelescope.org}}\cite{Doeleman2008,Akiyama2015,Fish2016, Goddi2017}. 
This imaging is performed by combining several radio telescopes into a
synchronised, global, and near Earth-sized network using
very-long-baseline interferometry (VLBI). 
However, another important goal of the
EHTC is the search for new radio pulsars in the vicinity of Sgr~A*.
Pulsars provide an additional independent observational tool
to help improve the understanding of the properties of Sgr~A*
(e.g., its mass, spin and even geometry),
providing considerably stronger constraints than is possible with
event horizon-scale imaging alone.

It is well known that pulsars, \ie rapidly rotating neutron stars, 
are among the most precise and stable
``clocks'' in the Universe, providing regular radio signals that can be
used to test GR and, in principle, any
alternative theory of gravity (see \cite{Lorimer2012} for a comprehensive
discussion).
In particular, when present in a binary system containing
another neutron star or a white dwarf, these objects arguably represent the
most promising avenue through which to investigate and constrain
large classes of gravity theories \cite{Freire2012}. 

The case of a pulsar orbiting around a SMBH is particularly interesting
since one can in principle combine precision timing measurements
with measurements of geodesic motion around the BH, \ie in the
strong-field regime. 
This system configuration has proven thus far to be elusive,
and consequently intensive searches by observational surveys like
BlackHoleCam\footnote{See
\href{https://blackholecam.org}{blackholecam.org}} and EHTC
will prove extremely important in view of these detections.
Such timing measurements can contribute to fixing strict
ranges on the parameters of a given class of gravity theories 
and therefore facilitate the selection of viable theories without
imposing any {\it a priori} assumptions.

Several different gravity theories can explain the
same experimental data with almost the same accuracy \cite{Planck2016a,BICEP2014}. 
The case of dark
matter is paradigmatic: astrophysical effects related to the older
concept of ``{missing matter}'' by Zwicky \cite{Zwicky1933} can be
addressed quite well by modifying the matter sector as well as the
gravity sector inside the field equations (see \cite{Capozziello2012} for a
comprehensive review). This degeneracy could be removed in favour of one
of the two approaches by either discovering new particles or by selecting
some ``new'' gravitational effect that clearly identifies a modified
theory.  At the same time, it is most desirable that the new
gravitational effect is measured in a way
that does not rely, {\it a priori}, on the selection of a given theory (or class
of theories) of gravity.

In order to address this specific problem, several authors have presented
novel and general approaches which enable the BH spacetime to
be parametrized based on
specific perturbations of, or deviations from, the general-relativistic Kerr metric
\cite{Johannsen2011,LinBambi2015}.
However, in this study theories of gravity entirely distinct from
GR are investigated, and consequently the parametrisation of Rezzolla and Zhidenko (RZ)
\cite{Rezzolla2014} is employed.

The RZ parametrization is a general representation of
BH spacetimes in arbitrary metric theories of gravity.
In the case of
spherically symmetric spacetimes, the parametrization makes use of a
coordinate compactification in terms of a rapidly-convergent
continued-fraction expansion defined in the radial direction between the
event horizon and spatial infinity. A similar approach has also been employed
to describe axisymmetric solutions, where the radial expansion is
accompanied by an expansion in the polar direction and away from the
equatorial plane \cite{Konoplya2016a}. In this way, it is possible to
represent a given BH solution to very high accuracy with a small number
of free parameters (see \cite{Younsi2016,Mizuno2017} for some examples of the
application of this parametrization to describe BH shadows).

The focus of the present study is restricted to the spherically symmetric
case and general expressions for the dynamics of a
test-particle in general BH spacetimes, such as the motion
of a pulsar orbiting around the SMBH candidate Sgr~A*, are derived.
In particular, explicit general expressions for the advance of the periastron
and the orbital period at different orders of the
parametrization are provided.
Furthermore, periastron-advance formulae are also given for four
representative theories, namely the Reissner-Nordstr\"{o}m solution
in GR, and alternative theory of gravity solutions from
Einstein-Maxwell-Axion-Dilaton, Brans-Dicke, and $f(R)$ theories.

The paper is organised as follows. In Sec.~\ref{due}
the RZ parametrization \cite{Rezzolla2014} is briefly reviewed,
while Sec.~\ref{tre} describes and compares the parametrization of other
representative spherically symmetric BH solutions. 
Section~\ref{quattro} discusses
the basic properties of test-particle motion around a
spherically symmetric BH, outlining the derivation of the expressions for
the advance of the periastron, which are then presented in Sec.~\ref{cinque}.  
These expressions are then employed
in Sec.~\ref{sei} to numerically investigate the values of the expansion
parameters in the case of four $S$-stars and four particular
representative pulsar toy models around Sgr.~A*.
Sec.~\ref{sette} is devoted to the discussion and conclusions.

\section{Parametrization Framework}
\label{due}
In what follows the RZ parametrization
\cite{Rezzolla2014} for a generic spherically symmetric BH spacetime is briefly reviewed
and subsequently used to determine the dynamics of a test
particle\footnote{Note that the RZ parametrization
  \cite{Rezzolla2014} does not provide any information on the field
  equations and hence can only be used to describe the motion of a test
  particle, be it massive (\eg a pulsar around a SMBH) or massless (\eg
  photons constituting the shadow of a BH \cite{Younsi2016}).}, such as a
pulsar orbiting around Sgr~A*. 
Unless otherwise stated, geometrised units with $G=c=1$ are used, where
$G$ and $c$ are Newton's constant and the speed of light, respectively.

Given a spherically-symmetric spacetime with line element given by
\footnote{The metric coefficients are written differently to
\cite{Rezzolla2014}: the present $ N(r)$ and $B(r)$ are 
precisely $N(r)^{2}$ and $B(r)^{2}$, respectively,
in \cite{Rezzolla2014}.}\cite{Rezzolla2014}
\begin{eqnarray}
\label{eq:RZ_metric}
ds^2 &=& g_{\alpha\beta} dx^{\alpha}dx^{\beta} \nonumber \\
        &=& N(r)dt^2-\frac{B(r)}{N(r)}dr^2-r^2 d\Omega^2 \,,
\end{eqnarray}
where the signature $(+,-,-,-)$ is adopted.
The geodesic equations of motion are derived from the Lagrangian, which
may be written as
\begin{eqnarray}
2{\cal L} &=& g_{\alpha\beta}\frac{dx^\alpha}{d\tau}\frac{dx^\beta}{d\tau} \nonumber \\
              &=& N(r){\dot t}^2-\frac{B(r)}{N(r)}{\dot r}^2-r^2 {\dot \theta}^2-r^2\sin ^2\theta 
                      {\dot \varphi}^2 \,,
\label{lagrangian}
\end{eqnarray}
where $\dot{x}^\mu := dx^\mu/d\tau$ is the particle's four-velocity, $\tau$ is
the affine parameter along the geodesic (in this work, the proper time), 
$t$ is the coordinate time, and an overdot denotes
differentiation with respect to $\tau$. In the RZ parametrization, the
function $N(r)$ is then expressed as
\begin{equation}
\label{N2}
N(x) = x A(x) \,,
\end{equation}
where
\begin{equation}
A(x) > 0 \quad\mbox{for}\quad 0\leq x\leq1 \,,
\end{equation}
with
\begin{equation}
x := 1-\frac{r_0}{r}\,,
\label{x}
\end{equation}
so that $x=0$ is the position of the event horizon and $x=1$ corresponds
to spatial infinity. Furthermore, $A$ and $B$ may be expressed in terms
of the parameters $\epsilon$, $a_i$, and $b_i$ ($i\in [0,n]$, where $n$ 
is the expansion order), such that
\begin{subequations}
\begin{align}
A(x) &= 1-\epsilon (1-x)+(a_0-\epsilon)(1-x)^2+{\widetilde A}(x)(1-x)^3 \,, \label{asympfix_1}  \\
B(x) &= 1+b_0(1-x)+{\widetilde B}(x)(1-x)^2 \,, \label{asympfix_2}
\end{align}
\end{subequations}
where the functions ${\widetilde A}$ and ${\widetilde B}$ describe the
metric near the horizon (\ie $x \simeq 0$) and at spatial infinity
(\ie $x \simeq 1$). 
It is evident that the metric is finite in
both limits \cite{Rezzolla2014}. 

The functions \eqref{asympfix_1}--\eqref{asympfix_2} can then be
expanded via a Pad\'e approximation of continuous as
\begin{subequations}
\label{contfrac}
\begin{align}
\label{contfrac_1}
{\widetilde A}(x)=\frac{a_1}{\displaystyle 1+\frac{\displaystyle
    a_2x}{\displaystyle 1+\frac{\displaystyle a_3x}{\displaystyle
      1+\ldots}}} \,,\\
\nonumber \\
\label{contfrac_2}
{\widetilde B}(x)=\frac{b_1}{\displaystyle 1+\frac{\displaystyle
    b_2x}{\displaystyle 1+\frac{\displaystyle b_3x}{\displaystyle
      1+\ldots}}}\,,
\end{align}
\end{subequations}
where $a_1, a_2,\ldots, a_{n}$ and $b_1, b_2,\ldots, b_{n}$ are dimensionless
constants that can be fixed once the generalised metric
\eqref{eq:RZ_metric} is matched to a specific metric. Hereafter, to keep
expressions compact, all calculations will be performed up to third order
in the above expansion; already at this order the differences between the
matched metric and the exact metric are below $1\%$
\cite{Rezzolla2014,Younsi2016,Kokkotas2017}. 
Finally, the parameter $\epsilon$ in equations \eqref{asympfix_1}--\eqref{asympfix_2} 
measures the deviations of the position of the event horizon
in the general metric from the corresponding location in a Schwarzschild
spacetime, \ie
\begin{equation}
\epsilon = \frac{2M-r_0}{r_0} = - \left(1 - \frac{2M}{r_0}\right) \,.
\label{epsilon}
\end{equation}

\section{Applications of the RZ parametrization}  
\label{tre}

There are several BH solutions which differ from GR, and in
order to perform a fair and unbiased analysis of the framework presented
in this study, in what follows no one model is favoured above others,
even though in practice there are physical motivations to do so depending
on the astrophysical applications in mind. Failure to do so would require
repeating the analysis for any and all models, which would be
time-consuming and impractical. In the following the advantage of this
parameterized approach which enables one to mimic different BHs is
demonstrated. In particular, four different representative BH solutions are
chosen to illustrate this.

First, a well-known spherically-symmetric solution of
GR is considered, namely the Reissner-Nordstr\"{o}m BH, which is in itself
interesting because it contains an electric charge. The presence of this
electric charge implies that the Reissner-Nordstr\"om BH can be more
compact than a Schwarzschild BH of the same mass.
Second, scalar-tensor theories, such as Brans-Dicke theory
and $f(R)$ theories, which are a major focus of a large proportion of the
gravitational physics community since they represent simple deviations
from (or extensions of) GR are next considered, providing interesting examples of
modified gravity theories.
Finally, a solution containing a dilaton scalar field, \ie an
Einstein-Maxwell-Axion-Dilaton BH, is also investigated.

\subsection{Einsteinian gravity: Reissner-Nordstr\"{o}m}
\label{tre.uno}
The Reissner-Nordstr\"{o}m metric describes the geometry
of a spherically symmetric and charged BH
\cite{Chandrasekhar83}. The line element can then be written
in the form \eqref{eq:RZ_metric} with
\begin{equation}
\label{RN}
N\left(r\right) = 1-\frac{2M}{r}+\frac{r_{_{\rm Q}}^2}{r^2} \,, \quad
B\left(r\right) = 1 \,,
\end{equation}
and $r_{_{\rm Q}}$ is a characteristic lengthscale given by
\begin{eqnarray}
r_{_{\rm Q}}^2=\frac{G\,Q^2}{4\pi \varepsilon_0} c^4 \,,
\end{eqnarray}
where $\left(4\pi \varepsilon_0\right)^{-1}$ is the Coulomb force constant.
In the limit $Q\rightarrow 0$, one recovers the Schwarzschild
solution. The Reissner-Nordstr\"{o}m solution posesses two event
horizons, which are located at
\begin{eqnarray}
r_{0,_{\rm RN}} = M\pm\sqrt{M^2-r^2_Q} \,.
\end{eqnarray}
Combining the above relation with \eqref{epsilon}, one may write $\epsilon$ for 
the Reissner-Nordstr\"{o}m solution as
\begin{eqnarray}
\epsilon_{_{\rm RN}} = \frac{2 M}{M+\sqrt{M^2-r_{_{\rm Q}}^2}}-1 \,. 
\end{eqnarray}
Upon expanding the metric coefficients \eqref{RN} at spatial infinity,
comparison with the RZ parametrization yields, at zeroth order
\begin{equation}
\label{a0RN}
a_{0,_{\rm RN}} = \frac{r_{_{\rm Q}}^2}{\left[\sqrt{M^2-r_{_{\rm Q}}^2}+M\right]^2} \,, \quad
b_{0,_{\rm RN}} = 0 \,.
\end{equation}
Similarly, comparing the behaviour of the coefficients \eqref{RN} near the horizon,
for all expansion orders $i\ge 1$, the following may be deduced
\begin{equation}
a_{i,{\rm RN}} = 0 \,, \quad b_{i,{\rm RN}} = 0 \,.
\end{equation}
In other words, the Reissner-Nordstr\"{o}m metric is fully represented by
the RZ parametrization at the zeroth order of the expansion. 

\subsection{Alternative theories of gravity}
\label{tre.due}

Alternative theories of gravity consider, in general, further (minimally
or non-minimally coupled) scalar fields or higher-order curvature or
torsion invariants in the Hilbert-Einstein Lagrangian. For example, if
correction terms such as
$\phi^2 R$, $R^2, \ R_{\alpha\beta}R^{\alpha\beta},
\ R_{\alpha\beta\gamma\delta}R^{\alpha\beta\gamma\delta}$ 
and $R\Box R$ are incorporated in
the Lagrangian, they give rise to modified gravitational dynamics
\cite{Capozziello2011,Nojiri2011,Cai2016}.
In particular, these modifications have been introduced in order to alleviate problems at ultraviolet scales (\eg divergences in quantum field theory, the lack of
a self-consistent quantum gravity theory, etc.) and at infrared scales
(\eg the cosmological accelerated expansion dubbed the ``{dark energy
  problem}'' and the clustering properties of large-scale structure, dubbed
the ``{dark-matter problem}'' \cite{Capozziello2012}. 

In particular, it is desirable to be able to calculate astrophysically
observable quantities in a way that does not rely on making any
assumption as to a particular theory of gravity. In this context, the RZ
parametrization provides a general approach that is
independent of the assumptions pertaining to a given theory of gravity
and describes the properties of test-particle motion, \eg the advance of
the periastron, simply in terms of the coefficients $a_{i}$ and $b_{i}$.
To this end, and principally in order to provide working examples, in the
following subsections three different classes of BH
solution in alternative theories of gravity are considered, namely: Brans-Dicke
theory, $f(R)$ gravity, and Einstein-Maxwell-Axion-Dilaton gravity.

\subsubsection{Brans-Dicke theory}
\label{tre.due.uno}

Brans-Dicke (BD) theory is the most well-known of the scalar-tensor theories.
In BD theory the gravitational interaction is mediated by a scalar field.  The
gravitational coupling is no longer constant, and instead $1/G$ is
replaced by a scalar field $\phi$ which is non-minimally coupled to the
Ricci scalar $R$ \cite{Brans1961}. The original approach was developed in
order to deal with a theory which could be more {Machian} than GR
\cite{Brans1961}. 
The BD field equations contain
a kinetic parameter, $\omega$, termed the BD coupling
constant. This is a dimensionless constant related to the strength and
variability of the scalar field, and whose value can be chosen to fit
observations. Using Solar-system tests it is possible to impose lower and
upper bounds on the possible values for $\omega$. Stringent limits on
$\omega$ can also be achieved through a consideration of the dynamics of
binary pulsars and some BH solutions can be derived in the framework of
this theory \cite{Damour1988,Damour1998, Thorne1971, Kim1999}. 
Hereafter the following BD solution is considered
\begin{eqnarray}
ds^2 = A(r)^{m+1} dt^2 - A(r)^{n-1} dr^2 - r^2 A(r)^n d\Omega^2 \,,
\nonumber\\
\label{BD}
\end{eqnarray}
with $\displaystyle{A(r)=1-{2{\tilde r}}/{r}}$, where $\tilde r$, $m$
and $n$ are arbitrary constants. The scalar field is given by
\begin{equation}
\phi(r) = \phi_0 A(r)^{-(m+n)/2} \,,
\end{equation}
with $\phi_0$ a constant. It is important to emphasise that the parameter
$n$ has the role of scalar hair and that as soon as $n=0$ the no-hair
condition is restored. It is immediately clear that for $m=n=0$ the
Schwarzschild solution is recovered, with asymptotic flatness being
recovered for any value of $m$ and $n$. The BD
parameter is found from the following relation
\begin{eqnarray}
\omega = -2 \left( 1 + \frac{m - n - mn}{(m+n)^2} \right) \,.
\label{omega}
\end{eqnarray}
In this manner, for each assigned value of $m$ and $n$, a class of BH
solutions is obtained. The event horizon is given by
\begin{equation}
r_{0,{{\rm BD}}} = 2{\tilde r} \label{BDEH} \quad \forall \quad m-n+1>0 \,.
\end{equation}
The parameter ${\tilde r}$ may be identified with $k M$, where $M$ is the BH mass
and $k$ is an arbitrary constant, yielding different event horizons.
For example, if $k=1$, \eqref{BDEH} corresponds to the
Schwarzschild event horizon and therefore $\epsilon=0$, as in the case of GR.
Fixing $n=0$ equation \eqref{BD} reduces to
\begin{eqnarray}
ds^2 = A(r)^{m+1} dt^2 - A(r)^{-1} dr^2 - r^2 d\Omega^2 \,,
\label{BDm}
\end{eqnarray}
and it is this particular case that is hereafter considered.
As a result, $\epsilon$ in this particular BD metric can be expressed as
\begin{equation}
\epsilon_{{\rm BD}} =-\left(1-\frac{M}{\tilde r}\right)\,.
\end{equation}
In general, expanding $A(r)$ in \eqref{BD} at infinity
yields the following expressions for $a_0$ and $b_0$ 
in terms of BD theory as
\begin{equation}
\label{a0BD}
a_{0,{\rm BD}} = \frac{\omega +3 (\omega +2)^2
  \epsilon_{{\rm BD}}}{2 (\omega +2)^2} \,, \quad
b_{0,{{\rm BD}}} = 0 \,.
\end{equation}
Furthermore, expanding near the horizon yields the following relations for the
parameters in terms of the theory are obtained
\begin{equation}
a_{1,{{\rm BD}}} = \frac{(\omega -2) \omega +3 (\omega +2)^3
  \epsilon_{{\rm BD}} }{6 (\omega +2)^3} \,,
\end{equation}
with 
\begin{equation}
a_{2,{\rm BD}}=0\,,\qquad {\rm and} \qquad 
b_{1,{\rm BD}}=b_{2,{\rm BD}}=0\,.
\end{equation}
Therefore, BD theory is represented by the
RZ parametrization at the first order of the expansion. 

\subsubsection{$f(R)$ gravity}
\label{tre.due.tre}

The general class of $f(R)$ theories relax the hypothesis that the
Hilbert-Einstein action must be linear in the Ricci scalar, and
instead assume general functions of $R$ that may be constrained by
observations and through theoretical considerations
\cite{Capozziello2008}. Such theories can always be reduced to
scalar-tensor theories by conformal transformations
\cite{Capozziello2011}. Due to this property, the above scheme can also be
adopted here and, in particular, the parametrized post-Newtonian
(PPN)\footnote{The post-Newtonian (PN) approach is an analytical
  approximation to GR based on a power series expansion in terms
  of the ratio $v/c$, where $v$ is the typical velocity of the system. In
  the limit where $v$ becomes infinite, the PN expansion reduces to
  Newton's law of gravity. The PPN parametrization uses the PN expansion
  to explicitly detail the parameters in which a general theory of
  gravity can differ from Newtonian gravity \cite{Will92}.}
parametrization arising from $f(R)$ gravity can be straightforwardly
related to the RZ parametrization. Assuming a static and
spherically symmetric metric, a general Post Newtonian (PN) approximation
can be written as \cite{Eddington1923}
\begin{equation}
\label{PNgen}
ds^2= N\left( r \right)dt^2 - \left( 1+\gamma\frac{2M}{r} \right) dr^2 -
r^2 d\Omega^2 \,,
\end{equation}
where
\begin{equation}
N\left( r \right) = 1-\frac{2M}{r}+\frac{\beta-\gamma}{2} \left( \frac{2M}{r}\right)^2 \,.
\end{equation}
The parameters $\gamma$ and $\beta$ provide a measure of the degree of
curvature of spacetime as generated by a body of mass $M$ at radius
$r$. Equation~\eqref{PNgen} is general and is valid for any metric theory
within which it is possible to derive the PN limit. This means that
metric coefficients and subsequently PN parameters strictly depend on the
choice of theory. The RZ parametrization can serve as a useful approach
in selecting viable $f(R)$ models within the PN-approximation.
Generalised PN-parameters can then be expressed in terms of $f(R)$
theory as \cite{Capozziello2005, Capozziello2006, Capozziello2009}
\begin{eqnarray}
\gamma_{\rm f(R)}  &=& 1 - \frac{\left[f''(R)\right]^2}{f'(R)+2\left[f''(R)\right]^2} \,, \label{gammaR} \\
\beta_{\rm f(R)} &=& 1 - \frac{1}{4} \left( \frac{f'(R)\, f''(R)}{2 f'(R)+3\left[f''(R)\right]^2} \right)
\frac{d\gamma_{\rm f(R)}}{dR} \,. \label{betaR}
\end{eqnarray}
It is evident that $\gamma_{\rm f(R)}$ and $\beta_{\rm f(R)}$ are strictly
dependent on the function $f(R)$ and its derivatives.
Here $f'(R):=df(R)/dR$. It is straightforward to demonstrate that for
$R\rightarrow\phi$, using a conformal transformation, one recovers
immediately the results for scalar tensor (ST) theories obtained in
\cite{Damour2007}:
\begin{eqnarray}
\label{gamma}
\gamma_{{\rm ST}} &=& 1 -\frac{\left[F'(\phi)\right]^2}{F(\phi)+2[F'(\phi)]^2}\,, \\ 
\beta_{{\rm ST}}
&=& 1 - \frac{1}{4} \left( \frac{F(\phi)\, F'(\phi)}{2F(\phi)+
  3[F'(\phi)]^2} \right) \frac{d\gamma}{d\phi}\,. \label{beta}
\end{eqnarray}
Here $\gamma_{{\rm ST}}$ and $\beta_{{\rm ST}}$
depend on the non-minimal coupling function $F(\phi)$ and its derivatives, and the
parameter $\alpha$ determines the deviation with respect to GR.
As a general consideration, it is
possible to say that both Eqs.~\eqref{gammaR}--\eqref{betaR} or
Eqs.~\eqref{gamma}--\eqref{beta} parameterize modified theories of
gravity according to a higher-order approach (\eg $f(R)$) or a
ST approach. The key point is that both pictures can be recast
in terms of the RZ parametrization and thus expressed in a general
approach which is effectively independent of the theory.

Comparing the expansions of the metrics \eqref{eq:RZ_metric} and
\eqref{PNgen} at the same order, it may be deduced that the event horizon for
a general $f(R)$ model is given by
\begin{equation}
r_{0,{{\rm f(R)}}} = M+M\sqrt{2 \gamma_{\rm f(R)}-2\beta_{\rm f(R)}+1} \,,
\end{equation}
where it is straightforward to see that when $f(R)=R$, the Schwarzschild
event horizon is readily recovered. Similarly, the
expression for $\epsilon$ in a $f(R)$ theory is found as
\begin{equation}
\epsilon_{\rm f(R)} = -\left(1 - \frac{2M}{r_{0,{{\rm f(R)}}}} \right) \,.
\end{equation}
Finally, asymptotically expanding the metrics \eqref{eq:RZ_metric} and 
\eqref{PNgen} and collecting the terms at equivalent expansion orders
yields the lowest-order expansion coefficients $a_{i,{{\rm f(R)}}}$ and
$b_{i,{{\rm f(R)}}}$ as
\begin{eqnarray}\label{a0fR}
a_{0,{{\rm f(R)}}} &=& \frac{(\beta_{\rm f(R)}-\gamma_{\rm f(R)})
                                                 (1+\epsilon_{\rm f(R)})^2}{2} \,, \\
b_{0,{{\rm f(R)}}} &=& \frac{(\gamma_{\rm f(R)}-1)
                                                 (1+\epsilon_{\rm f(R)})}{2} \,, \\
a_{1,{{\rm f(R)}}} &=& 3(a_{0,{\rm f(R)}}-\epsilon_{\rm f(R)})\,, \\
b_{1,{{\rm f(R)}}} &=& - 1 - b_{0,{\rm f(R)}} + 
                                                    \sqrt{1+\epsilon_{\rm f(R)}
                                                  + 2b_{0,{\rm f(R)}}}\,, \\
a_{2,{{\rm f(R)}}} &=& \frac{1}{a_{1,{\rm f(R)}}}
                                                  \left[3\left(\epsilon_{\rm f(R)}-1\right) - 
                                                  a_{0,{\rm f(R)}}\right]\,, \\
b_{2,{{\rm f(R)}}} &=& - 2+\frac{1}{b_{1,{\rm f(R)}}}
                                                    \left[\frac{1+\epsilon_{\rm f(R)} + 
                                                    2b_{0,{\rm f(R)}}}{2+b_{0,{\rm f(R)}}} - 
                                                    b_{0,{\rm f(R)}}\right] \,. \nonumber \\
\end{eqnarray}
As an example, a straightforward extension of any analytic $f(R)$
theory is the inclusion of a quadratic 
correction in the Ricci scalar,
the simplest correction to the standard Einstein-Hilbert
action (see \cite{Capozziello2012} for details on other and more complex
expressions for $f(R)$). In this case, $f(R)$ is simply expressed
as a Taylor series truncated at second order, \ie
\begin{equation} \label{fRL}
f(R) = R + \alpha\, R^2 + \ldots \,.
\end{equation}
Such a class of $f(R)$ theories has several applications, ranging from
Solar System scales up to early-Universe cosmology
\citep{DeLaurentis2015}. However, it is important to consider that the
range of values of $\alpha$ strictly depends on the scales under
consideration. For instance, it can be related to the scalaron or
inflaton mass and it must be compatible with the observed
amplitude of scalar perturbations, \ie in agreement with Planck
data \cite{Planck2016a,BICEP2014}.
In this work, however, the
length scales considered are much smaller than cosmological
scales, and so the resulting values for $\alpha$ are not those normally
adopted in the literature, 
\eg assuming dimensional units, a standard value for $\alpha$ can be $1/6$,
as obtained from conformal transformations \citep[see][for further details]{Birell82}. 
Nevertheless, theories of this type have been used recently
to describe gravitational corrections around SMBHs
\citep{Borka2016,Capozziello2014, Hees2017}.

\subsubsection{Einstein-Maxwell-Axion-Dilaton}
\label{tre.due.quattro}

The third alternative gravity theory considered in this work is the spherically
symmetric form of the Einstein-Maxwell-Axion-Dilaton (EMAD) gravity. 
In particular, the EMAD metric considered in this study is
spherically symmetric and is constructed from a simplification of the
axisymmetric EMAD solution \cite{Garcia1995} in the case of a vanishing
axion field. Solutions of this type arise from string theory
\cite{Gibbons1988, Garfinkle1991, Horowitz1991, Shapere1991,
Sen1992}. 
When the axion field vanishes and the BH is spherically symmetric, 
the EMAD BH is sometimes referred to as a ``dilaton'' BH 
and the line element takes the following form
\begin{equation} 
\label{EMAD}
ds^2 = \left(\frac{r-2\mu}{r+2\hat{b}}\right)dt^2
           - \left(\frac{r+2\hat{b}}{r-2\mu}\right)d\rho^2
           - (r^2+2\hat{b}r)d\Omega^2 \,,
\end{equation}
where
\begin{equation}
\mu := M - \hat{b} \,.
\end{equation}
Here $\hat{b}$ is the dilaton parameter and M the BH mass \citep[see][]{Sen1992}.
Upon recasting the radial coordinate as
\begin{equation}
\rho^2=r^2+2\hat{b}r\,,
\end{equation}
the EMAD line element may be re-expressed in terms of the RZ metric as
\begin{equation}
N\left( \rho \right) = 1-\frac{2M\rho}{r^2} \,, \quad B\left( \rho \right) = \frac{r^2}{\hat{b}^2+r^2} \,,
\end{equation}
where $r\equiv r(\rho)$.
Recalling that the location of the event horizon for this BH is given by
\begin{equation}
\rho_{0,_{{\rm DIL}}}= 2\left( M - \hat{b} \right) \,,
\end{equation}
and using Eq.~\eqref{epsilon}, the expression for $\epsilon$ in terms of
the axion-dilaton parameters may be written as
\begin{equation}
\epsilon_{_{\rm DIL}} = \sqrt{1+\frac{\hat{b}}{\mu}}-1 \,,
\end{equation}
where the subscript ``DIL'' refers to the dilaton BH.
In a similar manner, expanding the metric coefficients at spatial
infinity gives the values for $a_0$ and $b_0$ as
\begin{eqnarray}
\label{a0EMAD}
a_{0,_{{\rm DIL}}} = \frac{\hat{b}}{2\mu}\,, \quad {\text{ \rm and }}
                                              \quad b_{0,_{{\rm DIL}}}=0 \,.
\end{eqnarray}
In order to obtain the remaining coefficients, one must instead compare
the near-horizon expansions, obtaining
\begin{eqnarray}
a_{1,{_{\rm DIL}}} &=& - 3 - a_{0,{_{\rm DIL}}} + 2\left(\epsilon_{_{\rm DIL}}+1 \right) 
+ \left(1+a_{0,{_{\rm DIL}}}\right)^{-1} \,,
\nonumber \\ \\
b_{1,{_{\rm DIL}}} &=& \frac{\epsilon_{_{\rm DIL}}+1}{1+a_{0,{_{\rm DIL}}}}-1 \,, \\
a_{2,{_{\rm DIL}}} &=& \frac{ 2\left( a_{0,{_{\rm DIL}}} \epsilon_{_{\rm DIL}} - 
                                       a_{0,{_{\rm DIL}}}^2 +\epsilon_{_{\rm DIL}}\right) +1}
                                       {2 (a_{0,{_{\rm DIL}}}+1)^2} \,, \\
b_{2,{_{\rm DIL}}} &=& b_{1,{_{\rm DIL}}}-\frac{\hat{b}^2}{\left(1+a_{0,{_{\rm DIL}}}\right)^2} \,.
\end{eqnarray}
Note that the Schwarzschild BH solution is recovered from the dilaton BH
solution in the limit of vanishing dilaton parameter.

\section{Motion around a spherically symmetric black hole}
\label{quattro}

With the formalism derived so far, the
calculation of particle trajectories in the neighbourhood of 
a general spherically symmetric BH may now be calculated.
The canonical momenta may be expressed in terms
of the RZ parametrization as
\begin{eqnarray}
p_t &:=& \frac{\partial{\cal L}}{\partial\dot{t}} = + N(r) \ \! \dot{t} \,, \\
p_r &:=& \frac{\partial{\cal L}}{\partial\dot{r}} = - \frac{B(r)}{N(r)} \ \! \dot{r} \,, \\
p_\theta &:=& \frac{\partial{\cal L}}{\partial\dot{\theta}} = - r^2 \ \! \dot{\theta} \,,\\
p_\varphi &:=& \frac{\partial{\cal L}}{\partial\dot{\varphi}} = - r^2\sin^2\theta \ \! \dot{\varphi} \,.
\end{eqnarray}
Because of spherical symmetry, any orbital plane may be taken to be the
equatorial plane and therefore
$\theta=\pi/2$, $\dot \theta=\ddot \theta=0$ is assumed without loss of
generality. The integrals (constants) of motion may be written as
\begin{eqnarray}
p_t &=& N(r)\dfrac{dt}{d \tau} = E \,, \label{ptE} \\
p_\varphi &=& \hspace*{2mm} r^2 \dfrac{d \varphi}{d\tau} \hspace*{2mm}= -L \,, \label{pphiL}  
\end{eqnarray}
where $L$ denotes the component of the angular momentum of the particle
projected along the axis perpendicular to the orbital plane.

Furthermore, for motion in the equatorial plane, the total angular
momentum coincides with the azimuthal angular momentum.
Using eqs.~\eqref{ptE} and
\eqref{pphiL}, the Lagrangian can be rewritten as
\begin{equation}
2{\cal L} = \frac{E^2}{N(r)}-\frac{B(r)}{N(r)}{\dot r}^2-\frac{L^2}{r^2}= m^{2} \,.
\label{ep}
\end{equation}
where $m^{2}=(+1,0,-1)$ depending on whether the motion is timelike, null
or spacelike, respectively. Since the motion of a massive
particle around a SMBH is considered, $m^2=1$ is assumed hereafter. 
In this case, the constants of motion take the form
\begin{eqnarray}
E^2 &=& N(r) \left(\frac{L^2}{r^2}+1\right)+B(r) \ \! {\dot r}^2 \,, \label{energ} \\
L &=& r^2 \ \! {\dot \varphi} \,. \label{pphi}
\end{eqnarray}
The equation for the radial motion \eqref{energ} can then be written in
terms of an effective potential $V_{\rm eff}(r)$ in the following form
\begin{equation} 
\label{dotr1}
{\dot r}^2 = \frac{E^2}{B(r)}-\frac{N(r)}{B(r)}
\left(\frac{L^2}{r^2}+1\right) 
=\frac{E^2}{B(r)}-V^2_{\rm eff}(r) \,,
\end{equation}
where
\begin{equation} 
\label{eq:veff}
V^2_{\rm eff}(r) := -
\frac{N(r)}{B(r)} \left(\frac{L^2}{r^2}+1\right) \,.
\end{equation}

\section{Connecting to pulsar observations}
\label{cinque}

\subsection{Periastron advance and orbital period}
\label{cinque.uno}

As a direct application of the framework developed in the previous
sections, the transition between two close inner orbital turning
points (or, equivalently, between two close outer turning points) is
calculated. The orbits in this case remain closed if the magnitude
$\Delta \varphi$ of the angle swept out by the orbit is $2\pi$.
If this is not the case, then the inner turning points are
precessing and the amount of this precession per orbit is
\begin{equation} 
  \delta\varphi_{\rm prec} = \Delta \varphi - 2\pi \,. 
\end{equation}
In order derive the precession, $r$ must be expressed as a function of
$\varphi$ or vice-versa. Combining Eqs.~\eqref{pphi} and \eqref{dotr1},
one obtains
\begin{equation} \label{drdphi}
\left(\frac{dr}{d\varphi}\right)^2 = 
\frac{1}{L^{2}}\frac{N(r)}{B(r)} + 
\frac{1}{r^{2}}\frac{N(r)}{B(r)} 
- \frac{E^{2}}{L^{2}} \frac{1}{B(r)} \,.
\end{equation}
The angle $\Delta \varphi$ can be computed as 
the angle swept out as the particle passes between
the turning points $r_1$ and $r_2$ during its orbit, \ie
\begin{equation} \label{deltaphi}    
\Delta \varphi= \int^{r_2}_{r_1} dr \ \! \left(\frac{dr}{d\varphi}\right)^2 \,,
\end{equation}  
with the turning points $r_1$ and $r_2$ being determined from where
${\dot r}=0$ along the orbit. Using Eq.~\eqref{dotr1},
these points are determined by where the denominator of 
Eq.~\eqref{deltaphi} vanishes.

To illustrate how to derive a theory-independent expression for the
periastron advance within the RZ parametrization, first consider the
expansion of the metric when the only non-zero expansion terms are
$a_0$ and $b_0$.
Analysing Eq.~\eqref{drdphi} and remembering that the
bound (or unbound) nature of the orbits is determined by the energy $E$
and it is assumed that $E^2<1$, \ie the class of orbits considered herein
are characterised by a negative energy (\ie bound orbits). 
Due to the algebraic complexity
of the parametrization and the resulting equations of motion, only the
general expression is reported, and thus Eq.~\eqref{drdphi} may be written as
\begin{equation} 
\label{drdphi0}
\hspace{-7.5mm}{\phantom{\left(\frac{1}{1}\right)}}^{(0)}\!\!
                   \left(\frac{dr}{d\varphi}\right)^2 =
                   \frac{1}{{}^{(0)}L^2}\frac{{}^{(0)}N(r)}{ {}^{(0)}B(r)} + 
                   \frac{1}{r^2}\frac{{}^{(0)}N(r)}{{}^{(0)}B(r)}-\frac{{}^{(0)}E^2}
                     {{}^{(0)}L^2}\frac{1}{ {}^{(0)}B(r)} \,.
\end{equation}  
where ${}^{(n)}$ indicates working with all expansion coefficients up to
$a_n$ and $b_n$, \ie up to $n$-th order in the RZ parametrization and  ${}^{(0)}E$ and ${}^{(0)}L$ are the constants of motion related to the Lagrangian expanded to zeroth order. 
Upon changing variable as $u:=1/r$ and performing the necessary 
algebraic simplifications, one obtains
\begin{eqnarray} 
\label{dudphi0u}
\hspace{-7.5mm}{\phantom{\left(\frac{1}{1}\right)}}^{(0)}\!\!\left(\frac{du}{d\varphi}
                                     \right)^2 = &-& \frac{u (r_0 \epsilon+r_0)}{{}^{(0)}L^2} \nonumber \\
                 &+& \frac{u^2 \left[r_0^2 (a_0-\epsilon)+{}^{(0)}L^2+r_0^2 \epsilon\right]}
                           {{}^{(0)}L^2} \nonumber \\
             &-&\frac{u^3 \left[r_0^3 (a_0-\epsilon)+{}^{(0)}L^2 r_0 \epsilon + 
                     {}^{(0)}L^2 r_0\right]}{{}^{(0)}L^2}\nonumber \\
             &+& u^4r_0^2 a_0-r_0^3 u^5 (a_0-\epsilon) + 
                     \frac{1-{}^{(0)}E^2}{{}^{(0)}L^2}\,.
\nonumber \\
\end{eqnarray}
It is important to emphasize that in general all the $b_n$  terms disappear and
only the $a_n$ terms remain in the computation of the orbits.
Assuming $r_0 = 2M$ and
$a_0=\epsilon=0$, the well-known expression for the
Schwarzschild geodesic equations in terms of the variable $u$
are readily obtained
\cite{Landau-Lifshitz1, Roy2005}. Such expressions may be simplified
further by considering the following ansatz
\begin{equation} 
\label{orbita}
u = \frac{1 + e\cos \chi}{2\ell} \,,
\end{equation} 
where $e$ is the eccentricity and $\ell$ the {semi-latus rectum} of the
orbital ellipse, respectively. 
Here $\chi$ is the so-called {eccentric
  anomaly} or relativistic true anomaly
\cite{Darwin1961,Geisler1963,Chandrasekhar83, Frolov98, Maggiore2007}.
It is straightforward to express all the elements of orbits as integrals using
$\chi$ as the independent variable. According to Eq.~\eqref{orbita}, at
apoastron $\chi=\pi$, and at periastron $\chi=0$. It may be verified that
through several substitutions, Eq.~\eqref{drdphi0} reduces to the form
\begin{eqnarray} 
\label{dchidphi}
\hspace{-7.5mm}{\phantom{\left(\frac{1}{1}\right)}}^{(0)}
            \!\! \left(\frac{d\chi}{d\varphi}\right)^2&=&
            \left(\frac{d\chi}{d\varphi}\right)^2_{\rm GR} + \frac{1}{4}
            \left(1+e \cos\chi\right) \nonumber \\&&
            \times \left[ \epsilon - \sigma\left(a_{0}-\epsilon\right)\left(1+e \cos\chi\right) \right]
            \nonumber \\ &&
            \times \left\{ \left[2\left(e^2-4\right) \sigma^2+2\left(e^2+8\right)\sigma-8\right] \right. 
            \nonumber \\&&
           \ \ \ \ - \ e^3 \sigma^2 \cos\chi + 2e^2 \sigma(3\sigma-1) \cos2 \chi \nonumber \\&&
           \ \ \ \ + \ e^3 \sigma^2  \cos3\chi \left. \! \! \right\} \,,
\end{eqnarray}
where $\sigma:= r_{\rm g}/\ell$ and $r_{\rm g}$ is the gravitational
radius of the BH.
Here the general-relativistic contribution to the periastron
advance is given by \cite{Chandrasekhar83} 
\begin{equation} 
\left(\frac{d\chi}{d\varphi}\right)^2_{\rm GR} = 1 - \sigma \left(3+e\cos\chi\right) \,,
\end{equation}
which is immediately recovered from (\ref{dchidphi}) when
$a_0=\epsilon=0$. Upon integrating Eq.~\eqref{dchidphi}, considering the
semi-major axis $a$ of the orbital ellipse, and using the relation
$\ell=a(1-e^2)$, the expression for the periastron advance, $\Delta
\varphi$, at zeroth order in the RZ parametrization is obtained as
\begin{equation} 
\label{Prec0}
{}^{(0)} \Delta \widetilde{\varphi} = 
{3 \sigma}
+ 
\sigma \left(9 - 26 \sigma+20 \sigma^2 \right) a_0 +
20\sigma^{2} \left( 1 - \sigma \right) \epsilon \,,
\end{equation}
where $\Delta \widetilde{\varphi} := \Delta \varphi/2\pi$. The first
term in the above equation is simply the GR contribution, while the
remaining two terms represent the deviations from GR in a general BH spacetime
at zeroth order in the RZ parametrization.
 
Following the same procedure, after some simplification the periastron
advance may also be obtained for any order of the approximation. For
zeroth order, using Eq.~\eqref{drdphi} and expanding to first order, one
obtains upon integration the following expression
\begin{equation} \label{Prec1} 
  {}^{(1)}\Delta\widetilde{\varphi}= {}^{(0)}\Delta\widetilde{\varphi} + 
  4 \sigma^{2}\left(5-12\sigma+86\sigma^2\right) a_{1} \,.
\end{equation} 
Finally, the periastron advance may
be determined up to second order in the expansion. After the necessary
working this is obtained as
\begin{eqnarray} 
\label{Prec2}
{}^{(2)}\Delta\widetilde{\varphi} = {}^{(1)}\Delta\widetilde{\varphi}
&+& 2 \sigma \left(4 + \sigma \right) a_{2} \nonumber \\ 
&+& 2 \sigma \left(1 - 11\sigma + 28\sigma^2 - 20\sigma^3 \right) a_{0} a_{2} \nonumber \\
&+&   \sigma \left(1 + 14\sigma - 52\sigma^2 + 40\sigma^3 \right)a_{2} \epsilon \,. \nonumber \\
\end{eqnarray} 
The next step is to calculate the orbital period through
direct integration of eqs.~\eqref{ptE} and \eqref{pphiL} at the various
orders of the metric expansion \cite{Chandrasekhar83,Roy2005}.
In particular it is possible to distinguish the {anomalistic
period}, which is the time for the particle to travel from one
periastron to the next, and the {sidereal period}, which refers to the
lapse in time between two successive passages across a line through the
origin, fixed in space and lying in the orbital plane. For the sake of
brevity, only the expression for the zeroth order case is
given. 
The corresponding higher order expressions are
straightforward but algebraically cumbersome to write explicitly. Upon
using Eqs.~\eqref{orbita} and \eqref{dchidphi}, the proper time
of the test particle is obtained as
\begin{eqnarray} 
\label{tau}
{}^{(0)}\tau \ \ \!  = && \!\!\!\!\! \frac{p^{\frac{3}{2}}
  \sqrt{2 - \left(e^2+3\right) \sigma}}{\sqrt{r_0}}\int^{2\pi}_{\chi} {d\chi'}
\ \hspace{-7.5mm}{\phantom{\left(\frac{1}{1}\right)}}^{(0)}\!\!
\left(\frac{d\varphi}{d\chi'}\right) \nonumber \\ 
&\times& 
\frac{1}{\left(e \cos \chi'+1\right)^2 \left[1 - \sigma \left(e \cos \chi' +1\right)
    \right]} \,, \nonumber \\
\end{eqnarray}
and the coordinate time may then be expressed in terms of the
proper time as
\begin{equation} 
\label{t}
{}^{(0)} t = \sqrt{2} \left( \frac{\sqrt{\left( \sigma - 1 \right)^{2} - \sigma^{2}e^{2}}}
{\sqrt{2 - \left( e^{2} + 3 \right)\sigma}} \right) {}^{(0)}\tau\,.
\end{equation}  
These expressions result in an elliptic integral which must be solved
numerically. Expressing $t$ and $\tau$ in units of the Newtonian period
\begin{equation} 
\label{PNewton}
P_{\rm N}=\left( \frac{8\pi^2 a^3}{G r_0}\right)^{\frac{1}{2}} \,,
\end{equation}     
of a Keplerian orbit, the term multiplying the integral in
Eq.~\eqref{tau} is given by
\begin{equation} 
\frac{P_{\rm N}}{2\pi}(1-e^2)^{\frac{3}{2}}
\frac{\sqrt{2-\left(e^2+3\right)\sigma}}{\sqrt{2}} \,.
\end{equation}
The integrals of Eqs.~\eqref{tau} and \eqref{t} give $t$ and $\tau$ in
units of seconds upon restoring the proper physical unit values for $c$
and $G$. To calculate the above integrals numerical quadrature must be
used, since for a very eccentric orbit the integrand in \eqref{tau} can
be complex-valued. This can be avoided by re-expressing the integrals in
terms of the eccentric anomaly $\psi$, which is here related to $\chi$ as
\begin{equation}
\left(1 + e\cos\chi\right)\left(1 - e\cos\psi\right) = 1 - e^2 \,.
\end{equation}
This definition is chosen in analogy with the classical case
\cite{Damour1985,Damour1986}. It is clear from the above calculations
that one may obtain expressions for the orbital period in terms of the RZ
parametrization at any and all orders by determining only the correct
expression for $d\varphi / d\chi$.

\section{Application to astrophysical test cases}
\label{sei}
In the following sections both the
RZ parametrization and the expressions derived so far for the
periastron advance are tested. 
To do this, two sets of four test
objects are considered. 
The first set is represented by four well-known $S$-stars
which have now been observed orbiting
Sgr~A* for more than a decade \cite{Eckart1996, Eckart1997, Gillessen:2009, 
Gillessen2009L,Ghez1998, Ghez:2008}. 
These first objects are Keplerian but serve the purpose of providing representative
examples of how the parametrization can be employed. 
The second set of
test objects is represented by four pulsar toy models. 
Their properties have been chosen to have
a range of semi-major axes, reasonably high eccentricities and 
moderate-to-short orbital periods. 
In this respect, these toy models are idealised, but
not altogether unrealistic: future advances in instrumental sensitivity 
could, in principle, enable the detection of Galactic Centre pulsars with
such properties \cite{Lorimer2012}.

\begin{table*}
\centering
\caption{Values of the periastron advance $\Delta \varphi := 2\pi \, \Delta \widetilde{\varphi}$ for
different objects. Here the numerical values for GR displacement,
$\Delta\widetilde{\varphi}_{\rm GR}\,$ and for the RZ parametrization
\eqref{Prec0}-\eqref{Prec2} are shown.
The table reports the measured values of the eccentricity $e$,
semi-major axis $a$, and $\sigma=r_{\rm g}/\ell$, assuming that the
gravitational radius is $r_{\rm g}\sim6.64657 \times
10^9{\mathrm{m}}=0.044 \, \mathrm{AU}$ for Sgr~A*, and that the values 
of coefficients $a_0$, $a_1$, $a_2$ and $\epsilon$ are all set to
$10^{-3}$.
Numbers in square brackets denote multiplicative powers of ten.
}
\label{tab:1} 
\begin{tabular}{lccccccc}
\hline\hline
${\rm Object}\,$ &
$e\quad$ & $a \, [{\rm AU}] \quad $ & $\sigma$ &
$\Delta\widetilde{\varphi}_{\rm GR}$ & ${}^{(0)}\Delta\widetilde{\varphi}$ &  
${}^{(1)}\Delta\widetilde{\varphi}\,$ & ${}^{(2)}\Delta\widetilde{\varphi}$ \\
\hline
${S1}$ &
$0.358$& $3.29951 [+3]$ & $1.55089 [-5]$ &
$2.91131 [-4]$ & $2.92004 [-4]$ & $2.92005 [-4]$&
$2.92781 [-4]$ \\
\hline
${S2}$ &
$0.876$ & $9.79960 [+2]$ & $2.01195 [-4]$ &
$3.77681 [-3]$ & $3.78813 [-3]$ & $3.78814 [-3]$ &
$3.79821 [-3]$ \\
\hline
${S9}$ &
$0.825$ & $2.33559 [+3]$ & $6.01199 [-5]$ &
$1.12273 [-3]$ & $1.12610 [-3]$ & $1.12620 [-3]$ &
$1.12909 [-3]$ \\
\hline
${S13}$ &
$0.395$ & $9.53220 [+2]$ & $5.54552 [-5]$ &
$1.04410 [-3]$ & $1.04412 [-3]$ & $1.04413 [-3]$ &
$1.04690 [-3]$ \\
 \hline
${\rm Toy \ I}\quad$ &
$0.800$ & $1.75400 [+2]$ & $7.03608 [-4]$ &
$1.32627 [-2]$ & $1.33025 [-2]$ &
$1.33026 [-2]$ & $1.33379 [-2]$ \\
\hline
${\rm Toy \ II}\quad$ &
$0.800$ & $4.38500 [+1]$ & $2.81443 [-3]$ &
$5.30508 [-2]$ & $5.32097 [-2]$ &
$5.32107 [-2]$ & $5.33523 [-2]$ \\
\hline
${\rm Toy \ III}\quad$ &
$0.786$ & $ 5.00000 [+0]$ & $2.32488 [-2]$ &
$4.38229 [-1]$ & $4.39524 [-1]$ &
$4.39589 [-1]$ & $4.40764 [-1]$ \\
\hline
${\rm Toy \ IV}\quad$ & 
$0.888$ & $1.00000 [+0]$ & $2.10110 [-1]$ &
$3.96047 [+0]$ & $3.97069 [+0]$ &
$3.97765 [+0]$ & $3.98877 [+0]$ \\
\hline
\hline
\end{tabular}
\end{table*} 

\begin{figure*}
\includegraphics[width=0.99\columnwidth]{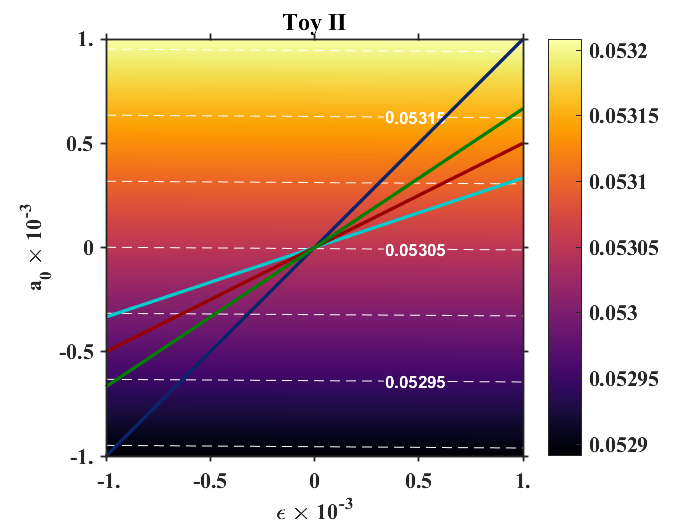} 
\includegraphics[width=0.99\columnwidth]{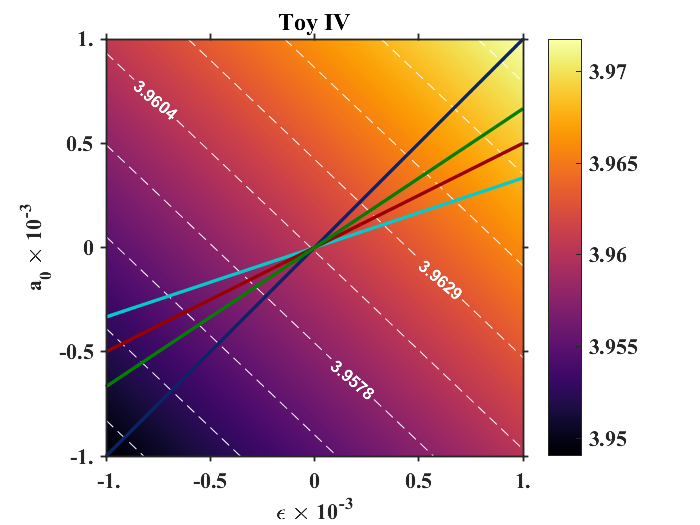} 
\caption{Contour plot of periastron advance for the toy models II and IV
  with contour lines indicating the value of the periastron advance for a
  given value of $a_0$ and $\epsilon$. Overlapping curves indicate the
  four different theories, which are rewritten in terms of the RZ
  parametrization, \ie \eqref{a0EMAD} (blue line), \eqref{a0RN} (green
  line), \eqref{a0fR} (red line), and \eqref{a0BD} (cyan
  line). \textit{Left panel}: Contour plot for toy model II. The
  separation between adjacent contours is
  $10^{-4}$. \textit{Right panel}: Contour plot for toy model IV. The
  separation between adjacent contours is $\sim 2.6 \times 10^{-3}$.}
\label{fig:1}
\end{figure*}

In modelling the toy pulsar-SMBH system it is hereafter assumed that
the mass of the central SMBH is known to some degree of precision. 
Whilst this is not the way pulsar timing normally works, 
since in such observations the mass of the BH is actually determined
from the observations of the binary system,
complementary observational data
(e.g.~multi-decadal observations of $S$-stars \cite{Boehle2016}), 
can provide an independent measurement of the BH mass.

\subsection{Determining the periastron advance and orbital period}
\label{sei.uno}
In the case of $S$-stars, the accuracy at which the advance of the
periastron can be measured places a lower limit on the eccentricity of
orbits that are in the range $0.35< e <0.93$.
For the purposes of this study four specific $S$-stars are considered,
namely $S1$, $S2$, $S9$, and $S13$, which represent a broad range in
both eccentricity and semi-major axis length 
\cite{Eckart1996, Eckart1997, Gillessen:2009, 
Gillessen2009L, Ghez1998, Ghez:2008}.

The properties of
these stars are collected in Table \ref{tab:1}, which also reports, besides
the eccentricities $e$ and semi-major axes $a$, the values of the
periastron advances $\Delta\widetilde{\varphi}_{\rm GR}$ and ${}^{(n)}\Delta\widetilde{\varphi}$
at the various orders, $n$, in the RZ expansion. Note that in evaluating the
periastron advance, specific values for the coefficients $\epsilon, a_i$ and $b_i$ must 
be specified, since these coefficients cannot yet be
constrained by astrophysical observations and 
hereafter chosen to be $a_0=a_1=a_2=\epsilon=10^{-3}$. 

Also reported in Table \ref{tab:1} are the values relative to the toy
pulsar models, where models I, II, III are in principle already measurable
with present radio-astronomical observations \citep[see][]{Liu2012},
while Toy IV is, at the present time, an optimistic model.
Since the RZ parametrisation is constructed to be most accurate at the
event horizon and at spatial infinity, the results presented in Table \ref{tab:1}
are, albeit weakly, dependent on the specific form of the parametrization.
However, given its inherent rapid convergence properties, by second order
the RZ parametrisation everywhere represents the chosen metric theory of gravity
to an accuracy of better than $1\%$ \cite{Rezzolla2014}.
In particular, looking at Table \ref{tab:1} one can
establish how well the RZ parametrization works in the vicinity of the
event horizon (this is especially true for Toy IV). In the case
of the $S$-stars, on the other hand, the reported deviations from GR are
all rather minute and this is to be expected since their motions are
essentially Keplerian.

Figure~\ref{fig:1} presents the results in Table \ref{tab:1},
highlighting how it is possible to use the results of the parametrization
to distinguish between different theories of gravity. 
In particular, the colour code in Fig.~\ref{fig:1} indicates the values of the
periastron advance at zeroth order, ${}^{(0)}\Delta\widetilde{\varphi}$, as a function of the
only two free parameters in the lowest-order expansion of the
parametrization, \ie $a_0$ and $\epsilon$. 
The left and right panels in Fig.~\ref{fig:1} refer to toy models II and IV, respectively.

In principle, from the observation of the periastron advance of a given star or pulsar,
the value of $\Delta \widetilde{\varphi}$ can be determined.
If the observation is performed over much longer time scales, e.g.~several decades
as in the case of $S$-stars, the accuracy of this measurement is significantly
improved.
The eccentricity and semi-major axis of the orbit are then determined,
and a contour plot akin to Fig.~\ref{fig:1} is made for the given object.
The observationally determined value of $\Delta \widetilde{\varphi}$ is then represented
as a contour line in this plot and deviations from GR 
(located at the point $(a_{0},\epsilon)=(0,0)$ in all plots) are manifest.
Different coloured lines indicate the constraints placed on $a_0$ and $\epsilon$ by the
different theories of gravity considered here, namely, Eq.~\eqref{a0EMAD}
(blue), Eq.~\eqref{a0RN} (green), Eq.~\eqref{a0fR} (red),
and Eq.~\eqref{a0BD} (cyan). 
Such lines may be constructed for any desired theory of gravity.
In principle, the intersection of the object's contour line with
any and all constructed (herein coloured) theory lines can provide
an estimate of both the deviation from GR and which potential
theories are more strongly (or weakly) constrained.
\begin{table*} 
\begin{center}
\caption{Values of the orbital period and its relative difference for
different objects. The Newtonian period ${\rm P_{Newton}}$, the GR
orbital period $\rm t_{GR}$ and the parameterised orbital period
${}^{(0)}t$, ${}^{(1)}t$ and ${}^{(2)}t$ are given, assuming that
$a_0=a_1=a_2=\epsilon=10^{-3}$. 
Numbers in square brackets denote multiplicative powers of ten.
}
\label{tab:2}
\begin{tabular}{lccccc}
\hline\hline
${\rm Object}$ & ${\rm P_{Newton} \, [s]}$ & ${\rm t_{GR} \, [s]}$ & 
${}^{(0)}t \, {\rm[s]}$ & ${}^{(1)}t \, {\rm[s]}$ & ${}^{(2)}t \, {\rm[s]}$ \\
\hline
${S1}$ & $2.81918 [+9]$ & $2.81924 [+9]$ & $2.82170 [+9]$ &
$2.82171 [+9]$ & $2.82298 [+9]$ \\
\hline
${S2}$ & $4.37764 [+8]$ & $4.37794 [+8]$ & $4.37896 [+8]$ &
$4.37896 [+8]$ & $4.39194 [+8]$ \\
\hline
${S9}$ & $1.67898 [+9]$& $1.67903 [+9]$ & $1.67956 [+9]$ &
$1.67956 [+9]$ & $1.68392 [+9]$ \\
\hline
${S13}$ & $4.37764 [+8]$ & $4.37794 [+8]$ & $4.38164 [+8]$ &
$4.38164 [+8]$ & $4.38408 [+8]$ \\
\hline
${\rm Toy \ I}$ & $3.45546 [+7]$ & $3.45678 [+7]$ & $3.45802 [+7]$ &
$3.45802 [+7]$ & $3.46642 [+7]$ \\ 
\hline
${\rm Toy \ II}$ & $4.31933 [+6]$ & $4.32593 [+6]$ & $4.32749 [+6]$ &
$4.32749 [+6]$ & $4.33803 [+6]$ \\
\hline
${\rm Toy \ III}$ & $1.66308 [+5]$ & $1.68621 [+5]$ & $1.68687 [+5]$ & $1.68687 [+5]$ & $1.69097 [+5]$ \\
\hline
${\rm Toy \ IV}$ & $1.48750 [+4]$ & $1.65483 [+4]$ & $1.65544 [+4]$ & $1.65539 [+4]$ & $1.66378 [+4]$ \\
\hline\hline
\end{tabular}
\end{center}
\end{table*} 
\begin{figure*}
\includegraphics[width=0.99\columnwidth]{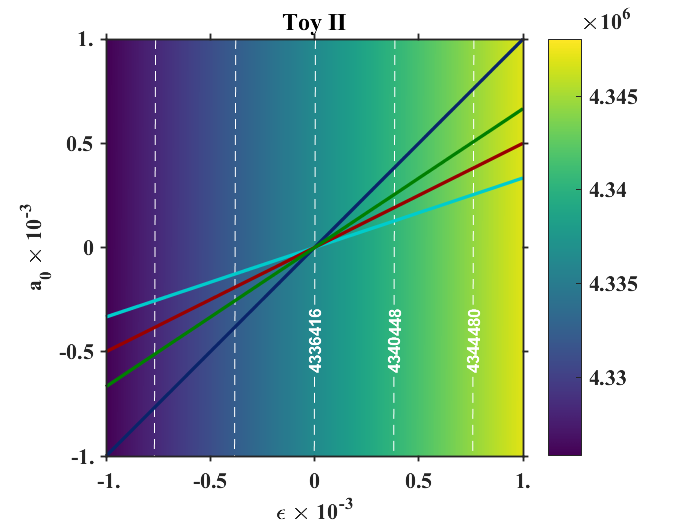}
\includegraphics[width=0.99\columnwidth]{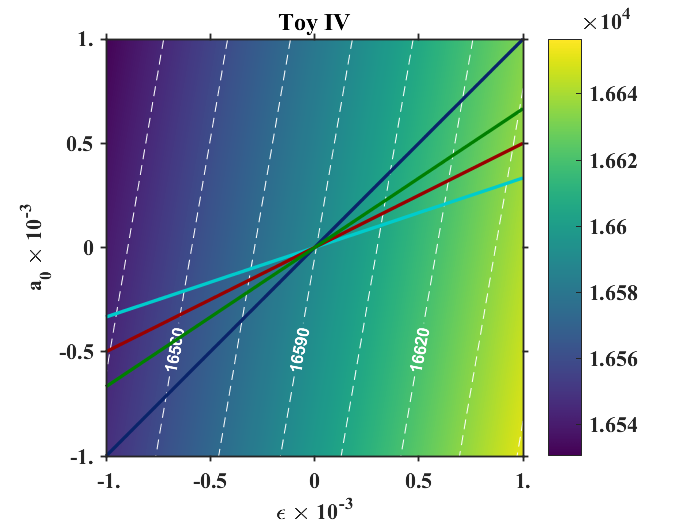}
\caption{Contour plot of the orbital period and curves as defined in
  Fig. \ref{fig:1}. \textit{Left panel}: Contour for the toy model
  II. The separation between adjacent contours is $4032$ s. \textit{Right
    panel}: Contour for the toy model IV. The separation between adjacent
  contours is $15$ s.}
\label{fig:2}
\end{figure*} 
Considerations similar to the ones made so far for the periastron advance
can also be made for the observed orbital period.
In particular, Table \ref{tab:2} and Fig.~\ref{fig:2} provide information
analogous to that presented in Table \ref{tab:1} and in
Fig.~\ref{fig:1}, but now for the orbital period. 
More specifically, Table \ref{tab:2} reports the orbital periods at different orders in the
expansion (see Sec.~\ref{sei.uno}). 
As expected, in the case of $S$-stars
the deviations from GR are very small since their motion is
effectively Keplerian, while larger deviations are seen for the pulsar
toy models and, in particular, for model IV. 
Table \ref{tab:2} also
demonstrates that the orbital period appears to be a more sensitive
parameter than the periastron advance 
(\ie deviations from GR are much more pronounced). 

Figure \ref{fig:2} presents a contour plot of
the values of the orbital period at the zeroth order in the
parametrization, ${}^{(0)}t$, as a function of the coefficients $a_0$ and
$\epsilon$. 
Analogous to Fig.~\ref{fig:1}, different coloured lines represent the
corresponding values for $a_0$ and $\epsilon$ predicted at this order by
the aforementioned different theories.

\begin{figure*}
\begin{center}
\includegraphics[width=0.40\textwidth]{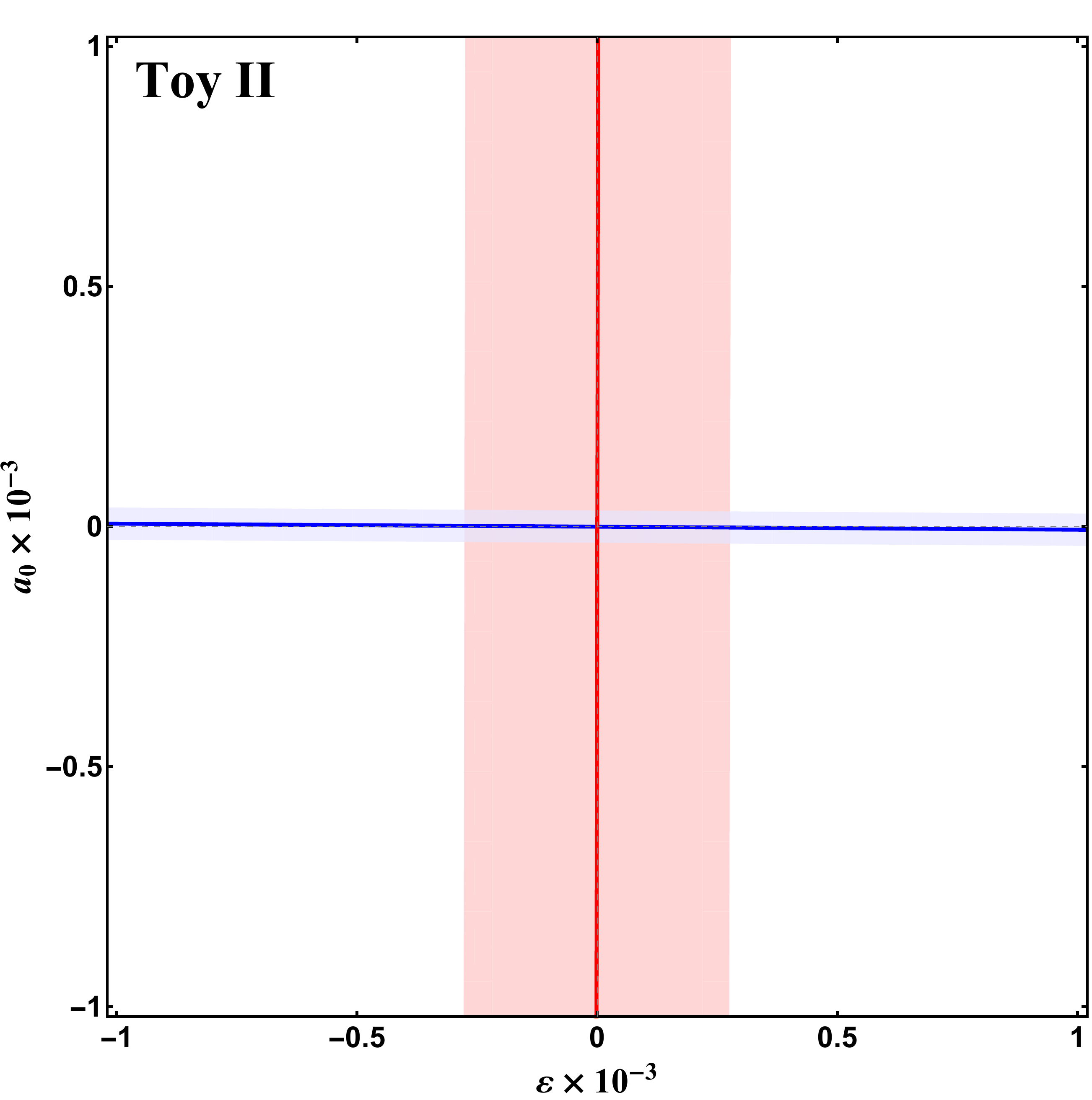}
\hspace{1.0cm}
\includegraphics[width=0.40\textwidth]{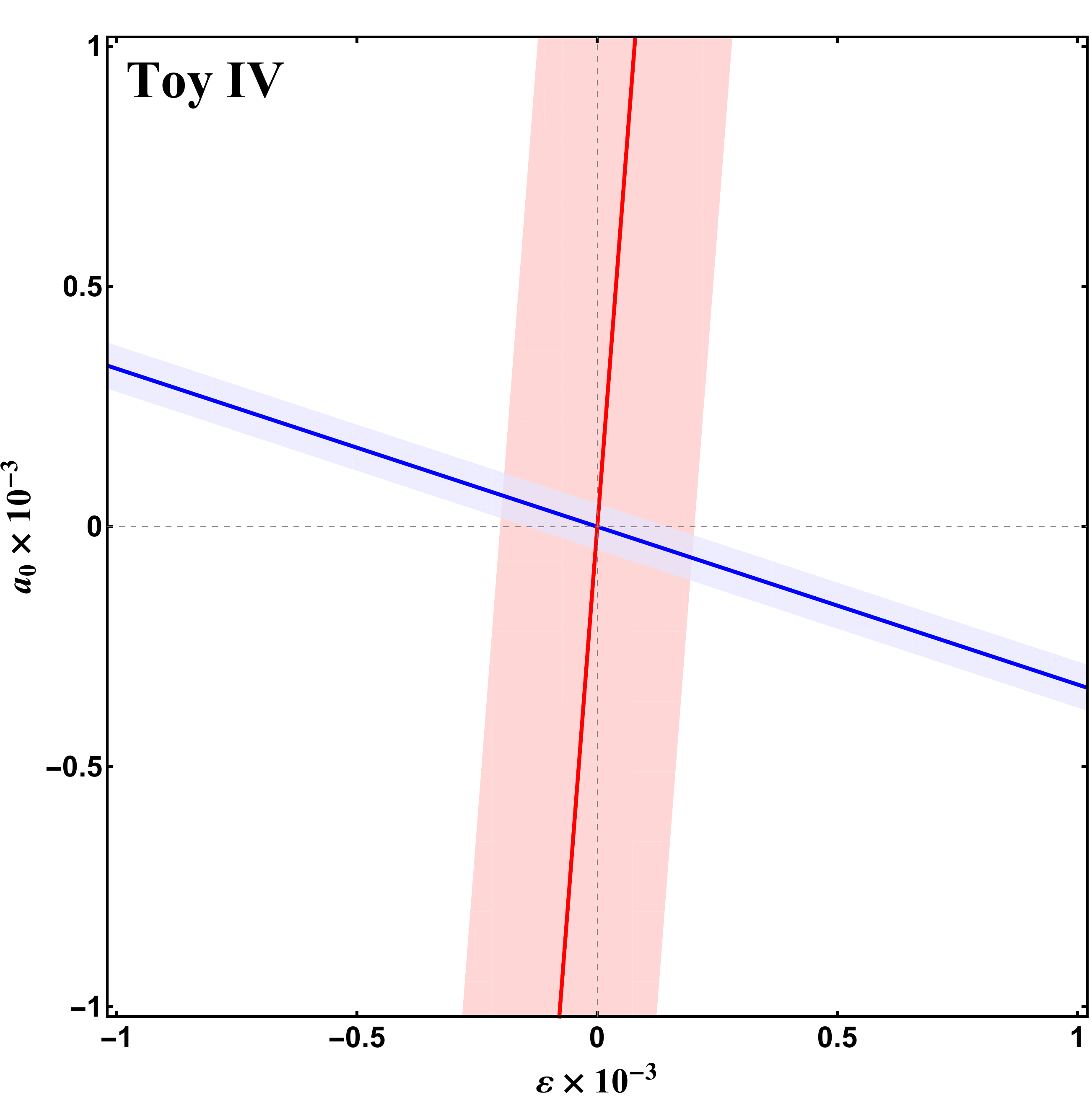}
\caption{Constraints set by the pulsar orbits for toy models II and IV in
the EMAD metric \eqref{EMAD} at zeroth order, with dilaton parameter
fixed at $\hat{b}=5 \times 10^{-8}$.
The values of the coefficients $a_0$ and $\epsilon$ are
constrained by the observations of the periastron advance 
(light-blue shaded areas around the solid blue line) and of the orbital period
(light-red shaded areas around the solid red line) 
assuming an accuracy of $10^{-4}$.
}
\label{fig:3}
\end{center}
\end{figure*}

As seen in Figs.~\ref{fig:1}--\ref{fig:2}, the allowed range of $a_{0}$ and $\epsilon$
is not yet bounded and thus (although increasingly unlikely for larger values of
$a_{0}$ and $\epsilon$) it is not possible to strongly constrain a particular theory.
However, simultaneous observation of both the periastron advance and orbital period 
of the object can (for a given accuracy) place constraints on the allowed range of
$a_{0}$ and $\epsilon$.
This in turn enables, upon re-examining Figs.~\ref{fig:1}--\ref{fig:2}, not only
much more stringent constraints to be imposed on given theories, but in
principle presents the possibility to rule out certain theories entirely.
This is illustrated in Fig.~\ref{fig:3} for the EMAD BH 
(assuming $\hat{b}=5 \times 10^{-8}$) using models Toy II 
and Toy IV, where the periastron advance (light-blue shaded areas around 
the solid blue line) and of the orbital period (light-red shaded areas around
the red line) assuming an accuracy of $10^{-4}$.
The specific case of a EMAD BH was chosen in this example since it is not
merely an extension of GR.

In summary, the results reported in
Figs.~\ref{fig:1}--\ref{fig:3} demonstrate that by using a
general description of test-particle motion in arbitrary BH spacetimes,
such as the RZ parameterization, future observations of pulsars
near a SMBH can help impose tighter constraints on 
different theories of gravity, and even potentially facilitate
ruling out certain theories 
(and, by extension, related classes and extensions thereof) entirely.

\subsection{Constraining EMAD and $f(R)$ theories}
\label{sei.tre}
%
In this subsection an illustrative example demonstrating the facility of the
parametric framework to constrain the parameters of two different
theories of gravity, namely EMAD and $f(R)$, is presented.
These two theories have been chosen since they are not only distinctly
different both mathematically and physically, but they also provide a strong contrast
in the ability of the RZ parametrization to constrain theory-dependent parameters.

\begin{figure*}
\includegraphics[width=0.43\textwidth]{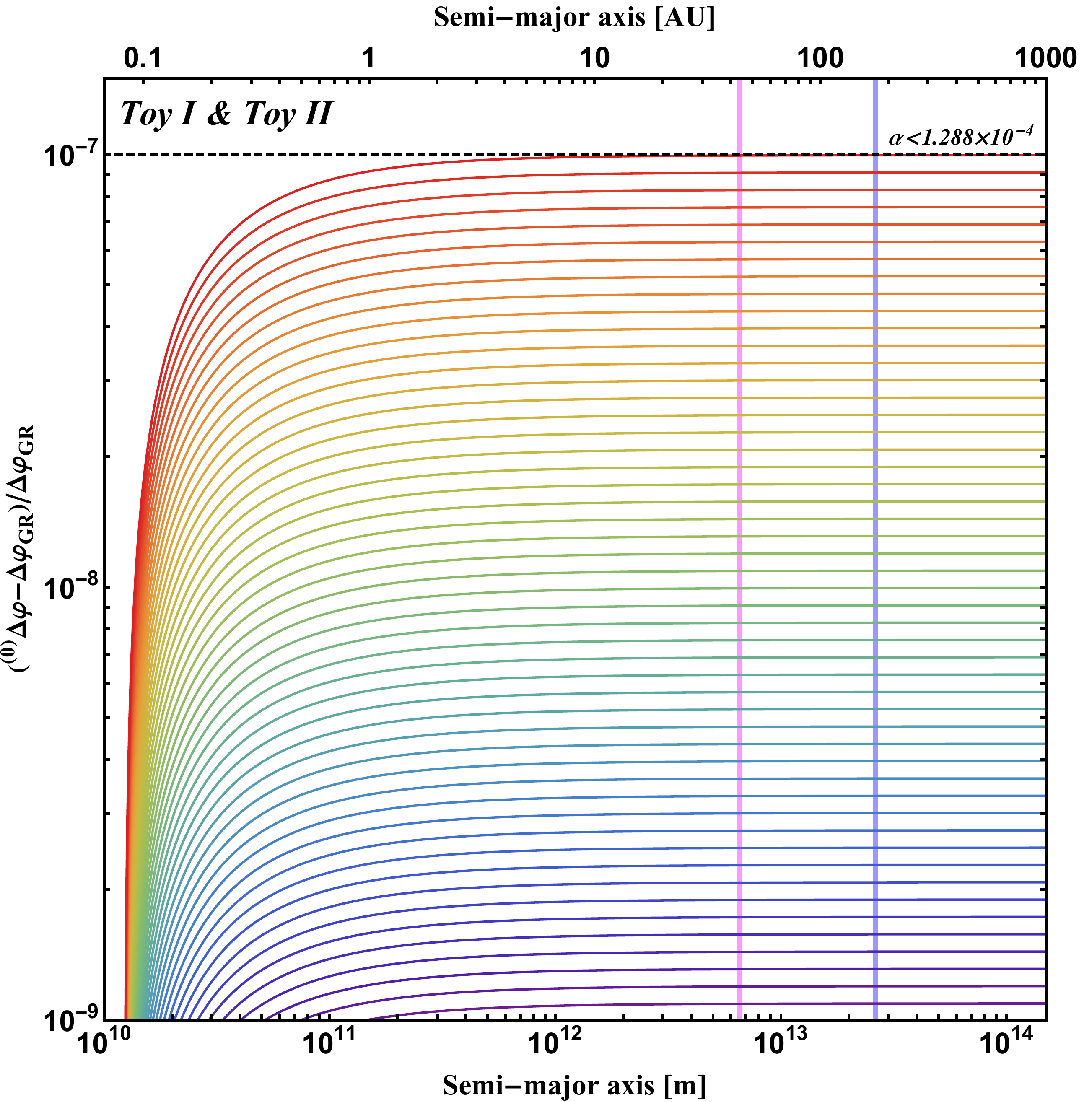} 
 \includegraphics[width=0.43\textwidth]{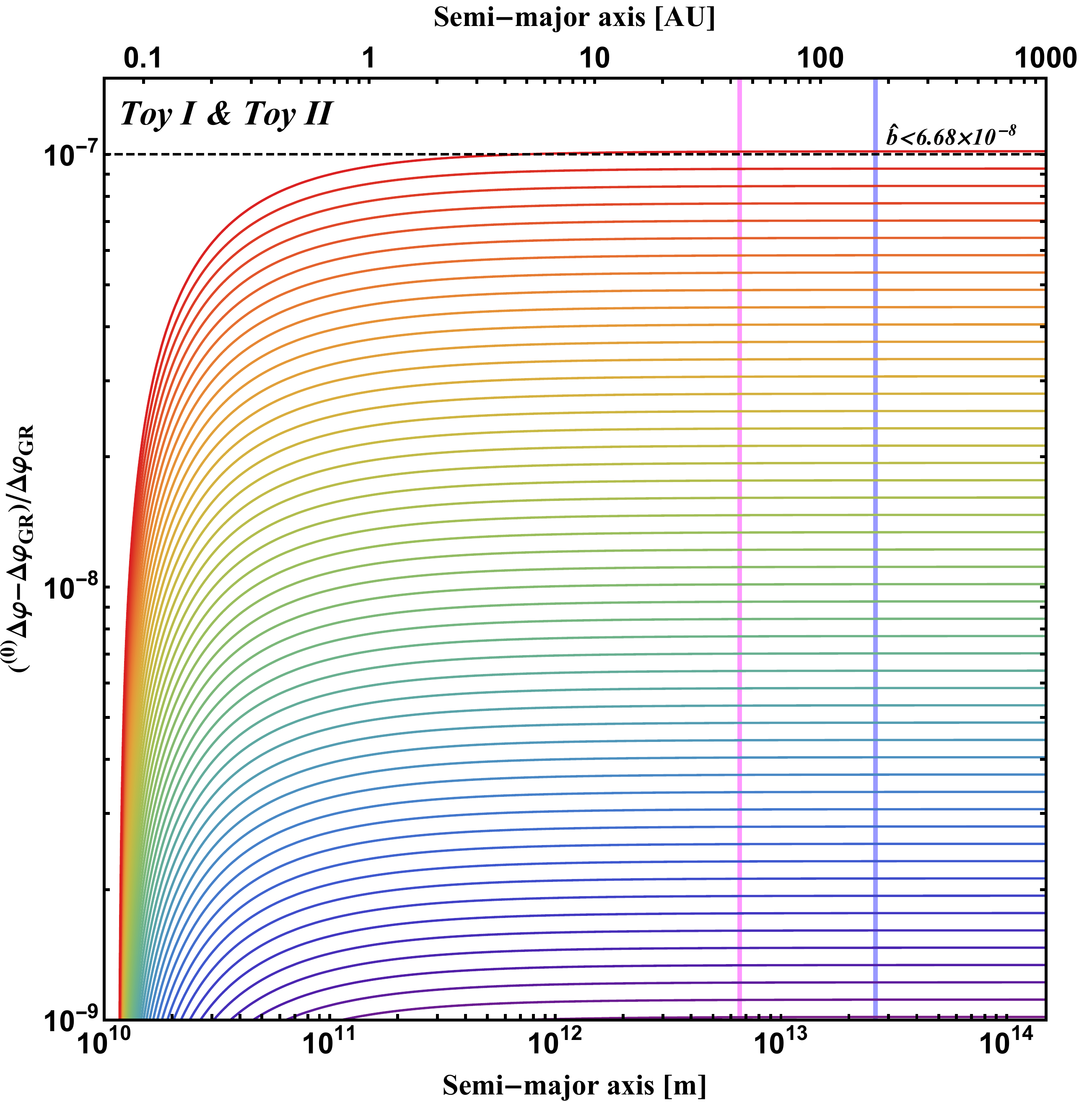} 
\caption{Relative difference of the zeroth order periastron advance with respect to
the GR value plotted as a function of semi-major axis distance for
Toy models I \& II in the case of $f(R)$ (left panel) and EMAD (right panel).
Different coloured lines indicate, for a fixed value of the theory parameter
($\alpha$ or $\hat{b}$), the variation in this relative difference as a function
of semi-major axis.
The 51 lines are equally spaced (logarithmically) between the stated inset 
upper limits of $\alpha \leq 1.288 \times 10^{-4}$ and $\hat{b} \leq 6.68 \times 10^{-8}$,
and 0.01 times those values, i.e. 25 equally spaced values per decade.
The verical solid lines at $175.4$ AU (blue) and $43.85$ AU (magenta) 
indicate the semi-major axis positions of Toy models I and II, respectively.
}
\label{fig:4}
\end{figure*}

\begin{figure*}
\includegraphics[width=0.43\textwidth]{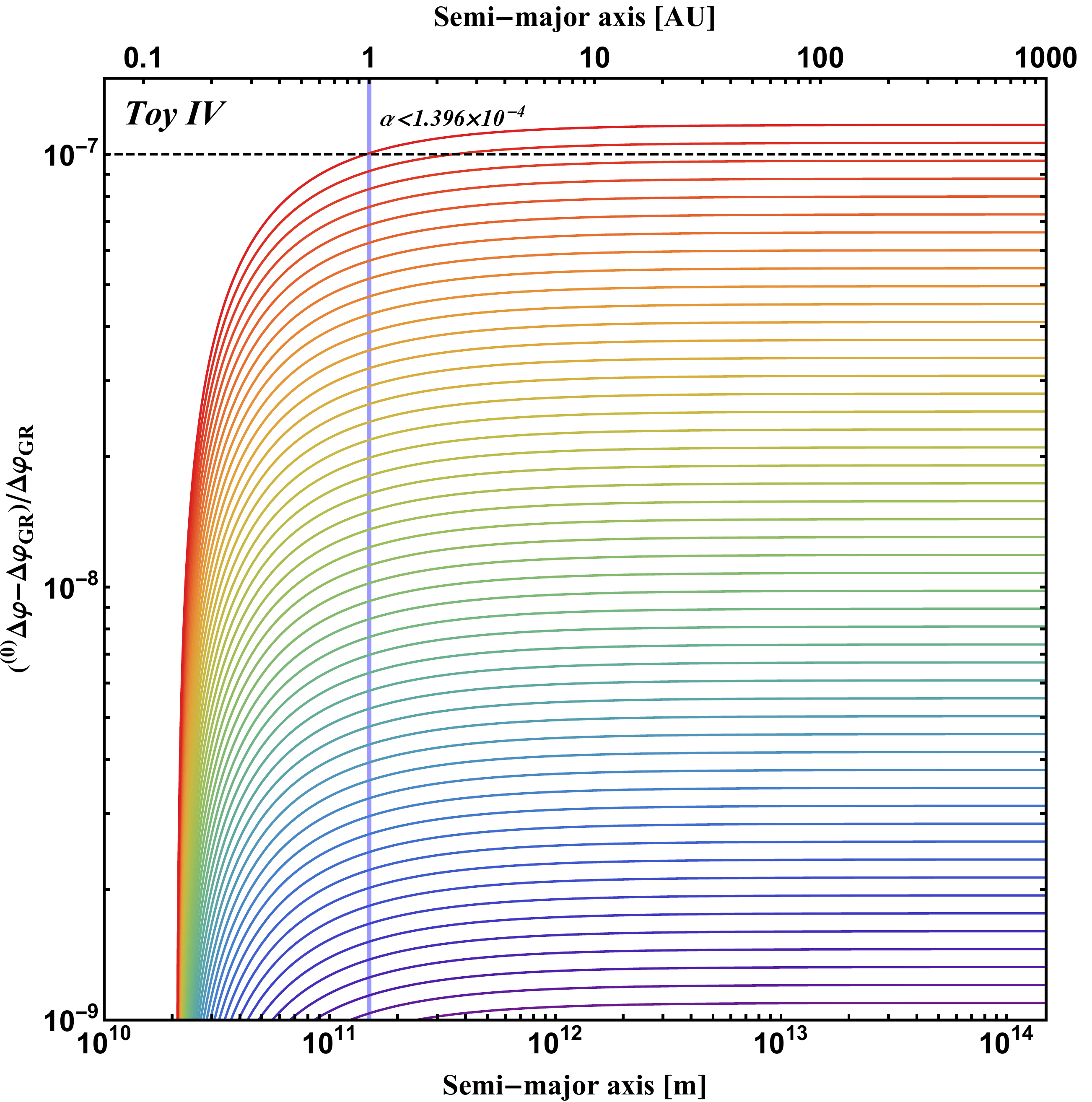} 
 \includegraphics[width=0.43\textwidth]{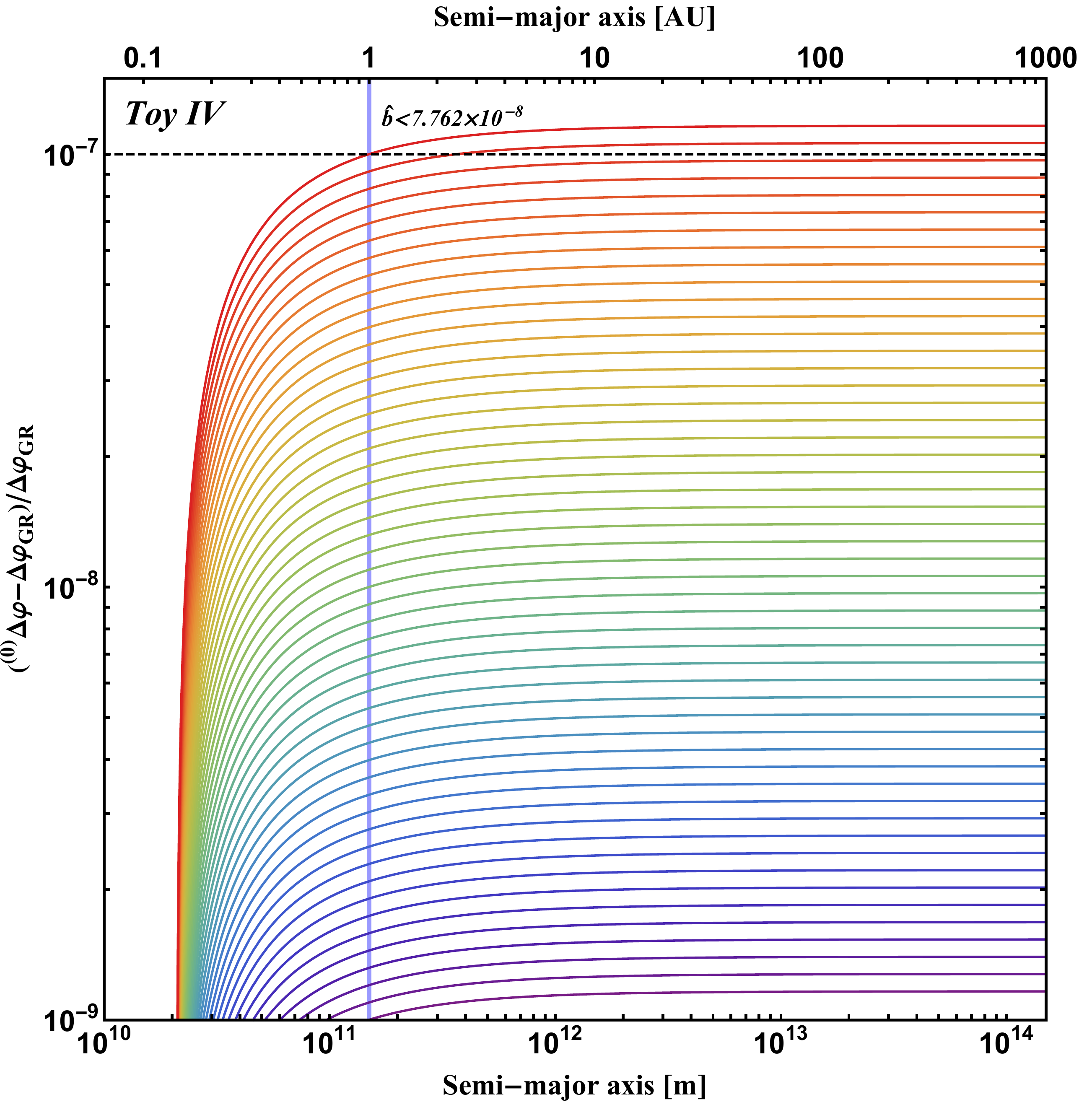}
\caption{Relative difference of the zeroth order periastron advance with respect to
the GR value plotted as a function of semi-major axis distance for
Toy models IV in the case of $f(R)$ (left panel) and EMAD (right panel).
Different coloured lines are as discussed in Fig.~\ref{fig:4},
with upper limits of $\alpha \leq 1.396 \times 10^{-4}$ and
$\hat{b} \leq 7.762 \times 10^{-8}$.
The vertical solid line (blue) at $1$ AU 
denotes the semi-major axis position of Toy model IV.
}
\label{fig:5}
\end{figure*}

Figures \ref{fig:4} and \ref{fig:5} present the deviations from GR of the periastron
advance as a function of the semi-major axis length, for models Toy I and Toy II
(Fig.~\ref{fig:4}) and model Toy IV (Fig.~\ref{fig:5}).
The left panels in both figures correspond to $f(R)$,
the right panels to EMAD.
Multiple coloured curves, coloured from violet through to red, denote,
for a fixed value of the theory parameter, how the relative difference in the
periastron advance varies as a function of the semi-major axis length.
These $51$ coloured lines are uniformly logarithmically spaced between the stated
parameter value in the upper right of each panel (uppermost red line) 
and $0.01$ times that value (bottommost violet line),
\ie $25$ lines per decade in the theory parameter.

In Fig.~\ref{fig:4} the relative differences in the periastron advance for models
Toy I and Toy II are shown for $f(R)$ (left panel), and EMAD (right panel).
The two vertical lines in both panels denote the semi-major axis position of
models Toy I (blue) and Toy II (magenta). 
The horizontal black dashed line at $10^{-7}$ represents a potential
astrophysical measurement precision \cite{Liu2012}.
It is immediately clear that for models Toy I and Toy II
the relative differences in the periastron advance
are insensitive to the semi-major axis length and practically indistinguishable,
as is evident from the horizontal, parallel theory parameter lines.
This can be interpreted as near-Keplerian pulsar motion, \ie the weak-field
limit of the RZ parametrization.
The intersection of the uppermost plotted theory parameter line with the vertical blue
and magenta lines provides an upper limit for the theory parameters.
For the $f(R)$ case this yields $\alpha < 1.288\times 10^{-4}$, whereas for the EMAD
case this yields the much smaller value of $\hat{b} < 6.68\times 10^{-8}$, nearly four
orders of magnitude smaller than the $f(R)$ case.

In Fig.~\ref{fig:5} the same analysis in Fig.~\ref{fig:4} is repeated for model Toy IV.
In this case the vertical blue line denotes the semi-major axis position of model Toy IV.
Since this model places the pulsar much closer to the event horizon of the BH ($1$~AU),
and as is clear from the theory parameter lines steeper gradients, the pulsar motion
can be considered as occurring in the transition region between the weak-field
and strong-field regimes.
For the $f(R)$ case this yields $\alpha < 1.396\times 10^{-4}$, whereas for the EMAD
case this yields $\hat{b} < 7.762\times 10^{-8}$.

In particular, one can see from Figs.~\ref{fig:4} and \ref{fig:5} that it is possible
to provide constraints on the two theories. More specifically, one can
place an upper limit on the value of the parameters $\alpha$ and $\hat{b}$.
It is assumed that the error on the measurements is of the order of $10^{-7}$
\cite{Liu2012}. Figures~\ref{fig:4} and \ref{fig:5} present the relative
differences from ${}^{(0)}\Delta\widetilde{\varphi}\,$ at first order of the
expansion in GR with respect to the semi-major axes for the four Toy
models for both theories. What emerges from the figures is that the
deviations from GR are much more remarkable if one considers, as seen
before, small semi-major axes and high eccentricities. In addition, one
may also estimate the various values of the parameters that characterise
the specific theory through the location of the different level curves.

It can be seen from Figs.~\ref{fig:4} and \ref{fig:5} that for a given measurement 
precision, as the pulsar semi-major axis location is shifted towards the BH
event horizon, the range over which the theory parameter can be probed 
(and effectively constrained) is increased.
Therefore, a pulsar orbiting in the immediate vicinity of a BH event horizon
(\ie in the truly strong-field regime) can provide much tighter constraints
on the theory parameters.

From Figs.~\ref{fig:4} and \ref{fig:5} it follows that $f(R)$ gravity
can be more stringently constrained using pulsars than EMAD theory.
This result is reassuring given that $f(R)$ is an extension
of GR, rather than a truly distinct theory of gravity like EMAD theory.
If a pulsar near Sgr~A*'s event horizon is detected, such observations will
aid in fixing ranges of validity and values of the parameters
of any theory under consideration. 
It is clear that in order to be proven useful, such a pulsar must be
close to the SMBH, with an orbital period of only a few hours or less.
Observational detection of such pulsars is in principle
achievable with present day radio telescopes.

\section{Conclusions} 
\label{sette}
Although GR has proven to be a reliable theory of gravity in several
different fields of application, and across several different scales,
it is by no means a complete one.
Furthermore, there exist other theories of gravity which
reproduce not only the results of classical GR, but also
the results of modern astrophysical observations (e.g.~in Cosmology) 
where pure GR fails to be as predictive.
There is therefore a pressing need, both theoretically and
observationally, to begin to, at the very least, impose constraints
(and even potentially exclude) particular theories of gravity.

Owing to their very narrow mass range, extreme compactness and
rapid, stable rotation periods, pulsars are one of the best candidates
to probe strong-field gravity in the truly nonlinear regime.
Pulsar-timing measurements, with their inherent high
precision, have already proven to be highly sensitive and accurate
in the weak-field regime \cite{Manchester2013}.
Measurement of a pulsar in the strong-field regime, \ie near a BH
event horizon, would enable not just a highly accurate determination of
the BH properties, but also provide an accurate probe of the spacetime
structure and geometry.

Given the rapid increase in recent efforts to perform astronomical
observations of the Galactic Center, the prospect of detecting a pulsar
orbiting in close proximity to Sgr~A* is promising.
Such a detection would provide the most accurate measurements of
the physical parameters (e.g.~mass, spin, and even quadrupole moment)
of Sgr~A* \cite{Psaltis2016}.
Theoretical studies of pulsar motion and timing in both GR and alternative
theories of gravity are therefore of great importance.
However, given the breadth of available theories of gravity in the present
literature, it is most expedient to perform any such studies in a manner
which assumes neither a specific theory of gravity nor any particular
solution to any particular theory.
It is also most desirable to have a representation in which the classical
GR limit is recovered.

Consequently, this study has presented an analysis of
test-particle (\ie pulsar) dynamics in several different BH spacetimes,
using a theory-independent approach.
This approach makes use of a general mathematical representation of
BH spacetimes.
In the case of spherically symmetric spacetimes,
\ie those considered in the present study,
a rapidly-convergent continued fraction expansion in terms of a
compactified radial coordinate has been employed.

Using this parametrization, general expressions for the
dynamics of a test-particle in general BH spacetimes were derived. 
In particular, general algebraic expressions for the advance
of the periastron and the orbital period at different orders of the
parametrization have been presented.
This formalism was applied to two sets of astrophysical test case:
(i) four particular $S$-stars, which have been observed orbiting Sgr~A*
in the infrared, and (ii) four hypothetical pulsar toy models.

It was shown that in the case of $S$-stars, deviations from GR are negligible
and therefore it is not possible to use such objects to test the underlying
theory of gravity.
The classical GR results of $S$-star periastron advance and orbital
period observations are well-reproduced within the paramterization and
it was shown that deviations from GR are negligible in this case.
Therefore $S$-stars, whilst useful in providing constraints on the mass
and distance of the central SMBH, are of limited use in probing
strong-field gravity.

Next, the periastron advance and orbital period properties of four pulsar toy
models were investigated.
It was shown that, in contrast to $S$-stars, pulsars with smaller semi-major
axis lengths (i.e. orbiting closer to Sgr~A*) indeed exhibit quantifiable
deviations from GR.
It was demonstrated that separate measurements of the periastron advance
and orbital period of the pulsar enable two independent means through which
to determine the parameters $a_{0}$ and $\epsilon$.
This presents the possibility to constrain any particular theory of gravity,
but due to the uncertainty in $a_{0}$ and $\epsilon$, does not
allow for the ruling-out of these theories. 
Whilst Solar System tests do impose some constraints, these are derived
in the weak field limit.

In order to attempt to rule out certain theories, an example of the simultaneous
calculation of the periastron advance and orbital period (complete with accuracy errors)
was presented.
This defines an inequality through which a region of overlap in the shared
parameter space of both observational quantities can be determined,
imposing tighter constraints on $a_{0}$ and $\epsilon$.
Such constraints can then be overlaid in Figs.~\ref{fig:1}--\ref{fig:2},
demonstrating that,in principle, given sufficient observational sensitivity,
certain theories of gravity (and even extensions or classes thereof) can
be immediately ruled out.

Finally, the particular cases of $f(R)$ and EMAD theories were employed to
show that measurements of the relative difference of the periastron advance
(from GR) can provide another avenue through which to constrain the parameters
of different theories of gravity.
Moreover, it was found that $f(R)$ theory is much more stringently constrained
than EMAD theory.
Given that the $f(R)$ gravity theory considered in this study is, at its core,
an extension of GR, this result, whilst not necessarily obvious, 
stands to reason given that this particular extension of GR includes
higher-order curvature invariants.
Therefore, using pulsar observations presents the possibility to strongly
constrain the parameters of all theories which are purely geometrical
(\ie not containing, \eg exotic particles and scalar fields).

In conclusion, pulsar observations can in principle accurately constrain the properties
of their central SMBH companion.
In this study, using a theory-independent parametric framework, it has
been shown that pulsar timing and dynamics also presents the possibility to
constrain (and even potentially exclude) theories of gravity.

\section*{Acknowledgements}

We thank Norbert Wex for numerous helpful discussions,
comments, and for helping construct the pulsar toy models.
We also thank C. Fromm, H. Olivares and A. Zhindenko for their
support and encouragement.
Support comes from the ERC Synergy Grant
``BlackHoleCam - Imaging the Event Horizon of Black Holes'' (Grant
610058), ``NewCompStar'', COST Action MP1304, the LOEWE-Program in
the Helmholtz International Center (HIC) for FAIR, and the European
Union's Horizon 2020 Research and Innovation Programme (Grant 671698)
(call FETHPC-1-2014, project ExaHyPE). 
Z.Y. acknowledges support from an Alexander von Humboldt Fellowship.

\bibliographystyle{apsrev4-1}
\bibliography{aeireferences}

\begin{thebibliography}{66}%
\makeatletter
\providecommand \@ifxundefined [1]{%
 \@ifx{#1\undefined}
}%
\providecommand \@ifnum [1]{%
 \ifnum #1\expandafter \@firstoftwo
 \else \expandafter \@secondoftwo
 \fi
}%
\providecommand \@ifx [1]{%
 \ifx #1\expandafter \@firstoftwo
 \else \expandafter \@secondoftwo
 \fi
}%
\providecommand \natexlab [1]{#1}%
\providecommand \enquote  [1]{``#1''}%
\providecommand \bibnamefont  [1]{#1}%
\providecommand \bibfnamefont [1]{#1}%
\providecommand \citenamefont [1]{#1}%
\providecommand \href@noop [0]{\@secondoftwo}%
\providecommand \href [0]{\begingroup \@sanitize@url \@href}%
\providecommand \@href[1]{\@@startlink{#1}\@@href}%
\providecommand \@@href[1]{\endgroup#1\@@endlink}%
\providecommand \@sanitize@url [0]{\catcode `\\12\catcode `\$12\catcode
  `\&12\catcode `\#12\catcode `\^12\catcode `\_12\catcode `\%12\relax}%
\providecommand \@@startlink[1]{}%
\providecommand \@@endlink[0]{}%
\providecommand \url  [0]{\begingroup\@sanitize@url \@url }%
\providecommand \@url [1]{\endgroup\@href {#1}{\urlprefix }}%
\providecommand \urlprefix  [0]{URL }%
\providecommand \Eprint [0]{\href }%
\providecommand \doibase [0]{http://dx.doi.org/}%
\providecommand \selectlanguage [0]{\@gobble}%
\providecommand \bibinfo  [0]{\@secondoftwo}%
\providecommand \bibfield  [0]{\@secondoftwo}%
\providecommand \translation [1]{[#1]}%
\providecommand \BibitemOpen [0]{}%
\providecommand \bibitemStop [0]{}%
\providecommand \bibitemNoStop [0]{.\EOS\space}%
\providecommand \EOS [0]{\spacefactor3000\relax}%
\providecommand \BibitemShut  [1]{\csname bibitem#1\endcsname}%
\let\auto@bib@innerbib\@empty
\bibitem [{\citenamefont {{Eckart}}\ and\ \citenamefont
  {{Genzel}}(1996)}]{Eckart1996}%
  \BibitemOpen
  \bibfield  {author} {\bibinfo {author} {\bibfnamefont {A.}~\bibnamefont
  {{Eckart}}}\ and\ \bibinfo {author} {\bibfnamefont {R.}~\bibnamefont
  {{Genzel}}},\ }\href {\doibase 10.1038/383415a0} {\bibfield  {journal}
  {\bibinfo  {journal} {Nature}\ }\textbf {\bibinfo {volume} {383}},\ \bibinfo
  {pages} {415} (\bibinfo {year} {1996})}\BibitemShut {NoStop}%
\bibitem [{\citenamefont {{Ghez}}\ \emph {et~al.}(2008)\citenamefont {{Ghez}},
  \citenamefont {{Salim}}, \citenamefont {{Weinberg}}, \citenamefont {{Lu}},
  \citenamefont {{Do}}, \citenamefont {{Dunn}}, \citenamefont {{Matthews}},
  \citenamefont {{Morris}}, \citenamefont {{Yelda}}, \citenamefont {{Becklin}},
  \citenamefont {{Kremenek}}, \citenamefont {{Milosavljevic}},\ and\
  \citenamefont {{Naiman}}}]{Ghez:2008}%
  \BibitemOpen
  \bibfield  {author} {\bibinfo {author} {\bibfnamefont {A.~M.}\ \bibnamefont
  {{Ghez}}}, \bibinfo {author} {\bibfnamefont {S.}~\bibnamefont {{Salim}}},
  \bibinfo {author} {\bibfnamefont {N.~N.}\ \bibnamefont {{Weinberg}}},
  \bibinfo {author} {\bibfnamefont {J.~R.}\ \bibnamefont {{Lu}}}, \bibinfo
  {author} {\bibfnamefont {T.}~\bibnamefont {{Do}}}, \bibinfo {author}
  {\bibfnamefont {J.~K.}\ \bibnamefont {{Dunn}}}, \bibinfo {author}
  {\bibfnamefont {K.}~\bibnamefont {{Matthews}}}, \bibinfo {author}
  {\bibfnamefont {M.~R.}\ \bibnamefont {{Morris}}}, \bibinfo {author}
  {\bibfnamefont {S.}~\bibnamefont {{Yelda}}}, \bibinfo {author} {\bibfnamefont
  {E.~E.}\ \bibnamefont {{Becklin}}}, \bibinfo {author} {\bibfnamefont
  {T.}~\bibnamefont {{Kremenek}}}, \bibinfo {author} {\bibfnamefont
  {M.}~\bibnamefont {{Milosavljevic}}}, \ and\ \bibinfo {author} {\bibfnamefont
  {J.}~\bibnamefont {{Naiman}}},\ }\href {\doibase 10.1086/592738} {\bibfield
  {journal} {\bibinfo  {journal} {Astrophys. J.}\ }\textbf {\bibinfo {volume}
  {689}},\ \bibinfo {pages} {1044} (\bibinfo {year} {2008})},\ \Eprint
  {http://arxiv.org/abs/0808.2870} {arXiv:0808.2870} \BibitemShut {NoStop}%
\bibitem [{\citenamefont {{Capozziello}}\ and\ \citenamefont {{de
  Laurentis}}(2011)}]{Capozziello2011}%
  \BibitemOpen
  \bibfield  {author} {\bibinfo {author} {\bibfnamefont {S.}~\bibnamefont
  {{Capozziello}}}\ and\ \bibinfo {author} {\bibfnamefont {M.}~\bibnamefont
  {{de Laurentis}}},\ }\href {\doibase 10.1016/j.physrep.2011.09.003}
  {\bibfield  {journal} {\bibinfo  {journal} {Phys. Rep.}\ }\textbf {\bibinfo
  {volume} {509}},\ \bibinfo {pages} {167} (\bibinfo {year} {2011})},\ \Eprint
  {http://arxiv.org/abs/1108.6266} {arXiv:1108.6266 [gr-qc]} \BibitemShut
  {NoStop}%
\bibitem [{\citenamefont {{Cunningham}}\ and\ \citenamefont
  {{Bardeen}}(1973)}]{Cunningham1973}%
  \BibitemOpen
  \bibfield  {author} {\bibinfo {author} {\bibfnamefont {C.~T.}\ \bibnamefont
  {{Cunningham}}}\ and\ \bibinfo {author} {\bibfnamefont {J.~M.}\ \bibnamefont
  {{Bardeen}}},\ }\href {\doibase 10.1086/152223} {\bibfield  {journal}
  {\bibinfo  {journal} {Astrophys. J.}\ }\textbf {\bibinfo {volume} {183}},\
  \bibinfo {pages} {237} (\bibinfo {year} {1973})}\BibitemShut {NoStop}%
\bibitem [{\citenamefont {{Falcke}}\ \emph {et~al.}(2000)\citenamefont
  {{Falcke}}, \citenamefont {{Melia}},\ and\ \citenamefont
  {{Agol}}}]{Falcke2000}%
  \BibitemOpen
  \bibfield  {author} {\bibinfo {author} {\bibfnamefont {H.}~\bibnamefont
  {{Falcke}}}, \bibinfo {author} {\bibfnamefont {F.}~\bibnamefont {{Melia}}}, \
  and\ \bibinfo {author} {\bibfnamefont {E.}~\bibnamefont {{Agol}}},\ }\href
  {\doibase 10.1086/312423} {\bibfield  {journal} {\bibinfo  {journal}
  {Astrophys. J. Lett.}\ }\textbf {\bibinfo {volume} {528}},\ \bibinfo {pages}
  {L13} (\bibinfo {year} {2000})},\ \Eprint
  {http://arxiv.org/abs/astro-ph/9912263} {astro-ph/9912263} \BibitemShut
  {NoStop}%
\bibitem [{\citenamefont {{Grenzebach}}\ \emph {et~al.}(2014)\citenamefont
  {{Grenzebach}}, \citenamefont {{Perlick}},\ and\ \citenamefont
  {{L{\"a}mmerzahl}}}]{Grenzebach14}%
  \BibitemOpen
  \bibfield  {author} {\bibinfo {author} {\bibfnamefont {A.}~\bibnamefont
  {{Grenzebach}}}, \bibinfo {author} {\bibfnamefont {V.}~\bibnamefont
  {{Perlick}}}, \ and\ \bibinfo {author} {\bibfnamefont {C.}~\bibnamefont
  {{L{\"a}mmerzahl}}},\ }\href {\doibase 10.1103/PhysRevD.89.124004} {\bibfield
   {journal} {\bibinfo  {journal} {Phys. Rev. D}\ }\textbf {\bibinfo {volume}
  {89}},\ \bibinfo {eid} {124004} (\bibinfo {year} {2014})},\ \Eprint
  {http://arxiv.org/abs/1403.5234} {arXiv:1403.5234 [gr-qc]} \BibitemShut
  {NoStop}%
\bibitem [{\citenamefont {{Abdujabbarov}}\ \emph {et~al.}(2015)\citenamefont
  {{Abdujabbarov}}, \citenamefont {{Rezzolla}},\ and\ \citenamefont
  {{Ahmedov}}}]{Abdujabbarov2015}%
  \BibitemOpen
  \bibfield  {author} {\bibinfo {author} {\bibfnamefont {A.~A.}\ \bibnamefont
  {{Abdujabbarov}}}, \bibinfo {author} {\bibfnamefont {L.}~\bibnamefont
  {{Rezzolla}}}, \ and\ \bibinfo {author} {\bibfnamefont {B.~J.}\ \bibnamefont
  {{Ahmedov}}},\ }\href {\doibase 10.1093/mnras/stv2079} {\bibfield  {journal}
  {\bibinfo  {journal} {Mon. Not. R. Astron. Soc.}\ }\textbf {\bibinfo {volume}
  {454}},\ \bibinfo {pages} {2423} (\bibinfo {year} {2015})},\ \Eprint
  {http://arxiv.org/abs/1503.09054} {arXiv:1503.09054 [gr-qc]} \BibitemShut
  {NoStop}%
\bibitem [{\citenamefont {{Younsi}}\ \emph {et~al.}(2016)\citenamefont
  {{Younsi}}, \citenamefont {{Zhidenko}}, \citenamefont {{Rezzolla}},
  \citenamefont {{Konoplya}},\ and\ \citenamefont {{Mizuno}}}]{Younsi2016}%
  \BibitemOpen
  \bibfield  {author} {\bibinfo {author} {\bibfnamefont {Z.}~\bibnamefont
  {{Younsi}}}, \bibinfo {author} {\bibfnamefont {A.}~\bibnamefont
  {{Zhidenko}}}, \bibinfo {author} {\bibfnamefont {L.}~\bibnamefont
  {{Rezzolla}}}, \bibinfo {author} {\bibfnamefont {R.}~\bibnamefont
  {{Konoplya}}}, \ and\ \bibinfo {author} {\bibfnamefont {Y.}~\bibnamefont
  {{Mizuno}}},\ }\href {\doibase 10.1103/PhysRevD.94.084025} {\bibfield
  {journal} {\bibinfo  {journal} {Phys. Rev. D}\ }\textbf {\bibinfo {volume}
  {94}},\ \bibinfo {eid} {084025} (\bibinfo {year} {2016})},\ \Eprint
  {http://arxiv.org/abs/1607.05767} {arXiv:1607.05767 [gr-qc]} \BibitemShut
  {NoStop}%
\bibitem [{\citenamefont {{Doeleman}}\ \emph {et~al.}(2008)\citenamefont
  {{Doeleman}}, \citenamefont {{Weintroub}}, \citenamefont {{Rogers}},
  \citenamefont {{Plambeck}}, \citenamefont {{Freund}}, \citenamefont
  {{Tilanus}}, \citenamefont {{Friberg}}, \citenamefont {{Ziurys}},
  \citenamefont {{Moran}}, \citenamefont {{Corey}}, \citenamefont {{Young}},
  \citenamefont {{Smythe}}, \citenamefont {{Titus}}, \citenamefont {{Marrone}},
  \citenamefont {{Cappallo}},\ and\ \citenamefont {et~al.}}]{Doeleman2008}%
  \BibitemOpen
  \bibfield  {author} {\bibinfo {author} {\bibfnamefont {S.~S.}\ \bibnamefont
  {{Doeleman}}}, \bibinfo {author} {\bibfnamefont {J.}~\bibnamefont
  {{Weintroub}}}, \bibinfo {author} {\bibfnamefont {A.~E.~E.}\ \bibnamefont
  {{Rogers}}}, \bibinfo {author} {\bibfnamefont {R.}~\bibnamefont
  {{Plambeck}}}, \bibinfo {author} {\bibfnamefont {R.}~\bibnamefont
  {{Freund}}}, \bibinfo {author} {\bibfnamefont {R.~P.~J.}\ \bibnamefont
  {{Tilanus}}}, \bibinfo {author} {\bibfnamefont {P.}~\bibnamefont
  {{Friberg}}}, \bibinfo {author} {\bibfnamefont {L.~M.}\ \bibnamefont
  {{Ziurys}}}, \bibinfo {author} {\bibfnamefont {J.~M.}\ \bibnamefont
  {{Moran}}}, \bibinfo {author} {\bibfnamefont {B.}~\bibnamefont {{Corey}}},
  \bibinfo {author} {\bibfnamefont {K.~H.}\ \bibnamefont {{Young}}}, \bibinfo
  {author} {\bibfnamefont {D.~L.}\ \bibnamefont {{Smythe}}}, \bibinfo {author}
  {\bibfnamefont {M.}~\bibnamefont {{Titus}}}, \bibinfo {author} {\bibfnamefont
  {D.~P.}\ \bibnamefont {{Marrone}}}, \bibinfo {author} {\bibfnamefont {R.~J.}\
  \bibnamefont {{Cappallo}}}, \ and\ \bibinfo {author} {\bibnamefont
  {et~al.}},\ }\href {\doibase 10.1038/nature07245} {\bibfield  {journal}
  {\bibinfo  {journal} {Nature}\ }\textbf {\bibinfo {volume} {455}},\ \bibinfo
  {pages} {78} (\bibinfo {year} {2008})},\ \Eprint
  {http://arxiv.org/abs/0809.2442} {arXiv:0809.2442} \BibitemShut {NoStop}%
\bibitem [{\citenamefont {{Akiyama}}\ \emph {et~al.}(2015)\citenamefont
  {{Akiyama}}, \citenamefont {{Lu}}, \citenamefont {{Fish}}, \citenamefont
  {{Doeleman}}, \citenamefont {{Broderick}}, \citenamefont {{Dexter}},
  \citenamefont {{Hada}}, \citenamefont {{Kino}}, \citenamefont {{Nagai}},
  \citenamefont {{Honma}}, \citenamefont {{Johnson}}, \citenamefont {{Algaba}},
  \citenamefont {{Asada}}, \citenamefont {{Brinkerink}}, \citenamefont
  {{Blundell}},\ and\ \citenamefont {et~al.}}]{Akiyama2015}%
  \BibitemOpen
  \bibfield  {author} {\bibinfo {author} {\bibfnamefont {K.}~\bibnamefont
  {{Akiyama}}}, \bibinfo {author} {\bibfnamefont {R.-S.}\ \bibnamefont {{Lu}}},
  \bibinfo {author} {\bibfnamefont {V.~L.}\ \bibnamefont {{Fish}}}, \bibinfo
  {author} {\bibfnamefont {S.~S.}\ \bibnamefont {{Doeleman}}}, \bibinfo
  {author} {\bibfnamefont {A.~E.}\ \bibnamefont {{Broderick}}}, \bibinfo
  {author} {\bibfnamefont {J.}~\bibnamefont {{Dexter}}}, \bibinfo {author}
  {\bibfnamefont {K.}~\bibnamefont {{Hada}}}, \bibinfo {author} {\bibfnamefont
  {M.}~\bibnamefont {{Kino}}}, \bibinfo {author} {\bibfnamefont
  {H.}~\bibnamefont {{Nagai}}}, \bibinfo {author} {\bibfnamefont
  {M.}~\bibnamefont {{Honma}}}, \bibinfo {author} {\bibfnamefont {M.~D.}\
  \bibnamefont {{Johnson}}}, \bibinfo {author} {\bibfnamefont {J.~C.}\
  \bibnamefont {{Algaba}}}, \bibinfo {author} {\bibfnamefont {K.}~\bibnamefont
  {{Asada}}}, \bibinfo {author} {\bibfnamefont {C.}~\bibnamefont
  {{Brinkerink}}}, \bibinfo {author} {\bibfnamefont {R.}~\bibnamefont
  {{Blundell}}}, \ and\ \bibinfo {author} {\bibnamefont {et~al.}},\ }\href
  {\doibase 10.1088/0004-637X/807/2/150} {\bibfield  {journal} {\bibinfo
  {journal} {Astrophys. J.}\ }\textbf {\bibinfo {volume} {807}},\ \bibinfo
  {eid} {150} (\bibinfo {year} {2015})},\ \Eprint
  {http://arxiv.org/abs/1505.03545} {arXiv:1505.03545 [astro-ph.HE]}
  \BibitemShut {NoStop}%
\bibitem [{\citenamefont {{Fish}}\ \emph {et~al.}(2016)\citenamefont {{Fish}},
  \citenamefont {{Johnson}}, \citenamefont {{Doeleman}}, \citenamefont
  {{Broderick}}, \citenamefont {{Psaltis}}, \citenamefont {{Lu}}, \citenamefont
  {{Akiyama}}, \citenamefont {{Alef}}, \citenamefont {{Algaba}}, \citenamefont
  {{Asada}}, \citenamefont {{Beaudoin}}, \citenamefont {{Bertarini}},
  \citenamefont {{Blackburn}}, \citenamefont {{Blundell}}, \citenamefont
  {{Bower}},\ and\ \citenamefont {et~al.}}]{Fish2016}%
  \BibitemOpen
  \bibfield  {author} {\bibinfo {author} {\bibfnamefont {V.~L.}\ \bibnamefont
  {{Fish}}}, \bibinfo {author} {\bibfnamefont {M.~D.}\ \bibnamefont
  {{Johnson}}}, \bibinfo {author} {\bibfnamefont {S.~S.}\ \bibnamefont
  {{Doeleman}}}, \bibinfo {author} {\bibfnamefont {A.~E.}\ \bibnamefont
  {{Broderick}}}, \bibinfo {author} {\bibfnamefont {D.}~\bibnamefont
  {{Psaltis}}}, \bibinfo {author} {\bibfnamefont {R.-S.}\ \bibnamefont {{Lu}}},
  \bibinfo {author} {\bibfnamefont {K.}~\bibnamefont {{Akiyama}}}, \bibinfo
  {author} {\bibfnamefont {W.}~\bibnamefont {{Alef}}}, \bibinfo {author}
  {\bibfnamefont {J.~C.}\ \bibnamefont {{Algaba}}}, \bibinfo {author}
  {\bibfnamefont {K.}~\bibnamefont {{Asada}}}, \bibinfo {author} {\bibfnamefont
  {C.}~\bibnamefont {{Beaudoin}}}, \bibinfo {author} {\bibfnamefont
  {A.}~\bibnamefont {{Bertarini}}}, \bibinfo {author} {\bibfnamefont
  {L.}~\bibnamefont {{Blackburn}}}, \bibinfo {author} {\bibfnamefont
  {R.}~\bibnamefont {{Blundell}}}, \bibinfo {author} {\bibfnamefont {G.~C.}\
  \bibnamefont {{Bower}}}, \ and\ \bibinfo {author} {\bibnamefont {et~al.}},\
  }\href {\doibase 10.3847/0004-637X/820/2/90} {\bibfield  {journal} {\bibinfo
  {journal} {Astrophys. J.}\ }\textbf {\bibinfo {volume} {820}},\ \bibinfo
  {eid} {90} (\bibinfo {year} {2016})},\ \Eprint
  {http://arxiv.org/abs/1602.05527} {arXiv:1602.05527} \BibitemShut {NoStop}%
\bibitem [{\citenamefont {{Goddi}}\ \emph {et~al.}(2017)\citenamefont
  {{Goddi}}, \citenamefont {{Falcke}}, \citenamefont {{Kramer}}, \citenamefont
  {{Rezzolla}}, \citenamefont {{Brinkerink}}, \citenamefont {{Bronzwaer}},
  \citenamefont {{Davelaar}}, \citenamefont {{Deane}}, \citenamefont {{de
  Laurentis}}, \citenamefont {{Desvignes}}, \citenamefont {{Eatough}},
  \citenamefont {{Eisenhauer}}, \citenamefont {{Fraga-Encinas}}, \citenamefont
  {{Fromm}}, \citenamefont {{Gillessen}},\ and\ \citenamefont
  {et~al.}}]{Goddi2017}%
  \BibitemOpen
  \bibfield  {author} {\bibinfo {author} {\bibfnamefont {C.}~\bibnamefont
  {{Goddi}}}, \bibinfo {author} {\bibfnamefont {H.}~\bibnamefont {{Falcke}}},
  \bibinfo {author} {\bibfnamefont {M.}~\bibnamefont {{Kramer}}}, \bibinfo
  {author} {\bibfnamefont {L.}~\bibnamefont {{Rezzolla}}}, \bibinfo {author}
  {\bibfnamefont {C.}~\bibnamefont {{Brinkerink}}}, \bibinfo {author}
  {\bibfnamefont {T.}~\bibnamefont {{Bronzwaer}}}, \bibinfo {author}
  {\bibfnamefont {J.~R.~J.}\ \bibnamefont {{Davelaar}}}, \bibinfo {author}
  {\bibfnamefont {R.}~\bibnamefont {{Deane}}}, \bibinfo {author} {\bibfnamefont
  {M.}~\bibnamefont {{de Laurentis}}}, \bibinfo {author} {\bibfnamefont
  {G.}~\bibnamefont {{Desvignes}}}, \bibinfo {author} {\bibfnamefont {R.~P.}\
  \bibnamefont {{Eatough}}}, \bibinfo {author} {\bibfnamefont {F.}~\bibnamefont
  {{Eisenhauer}}}, \bibinfo {author} {\bibfnamefont {R.}~\bibnamefont
  {{Fraga-Encinas}}}, \bibinfo {author} {\bibfnamefont {C.~M.}\ \bibnamefont
  {{Fromm}}}, \bibinfo {author} {\bibfnamefont {S.}~\bibnamefont
  {{Gillessen}}}, \ and\ \bibinfo {author} {\bibnamefont {et~al.}},\ }\href
  {\doibase 10.1142/S0218271817300014} {\bibfield  {journal} {\bibinfo
  {journal} {International Journal of Modern Physics D}\ }\textbf {\bibinfo
  {volume} {26}},\ \bibinfo {eid} {1730001-239} (\bibinfo {year} {2017})},\
  \Eprint {http://arxiv.org/abs/1606.08879} {arXiv:1606.08879 [astro-ph.HE]}
  \BibitemShut {NoStop}%
\bibitem [{\citenamefont {{Lorimer}}\ and\ \citenamefont
  {{Kramer}}(2012)}]{Lorimer2012}%
  \BibitemOpen
  \bibfield  {author} {\bibinfo {author} {\bibfnamefont {D.~R.}\ \bibnamefont
  {{Lorimer}}}\ and\ \bibinfo {author} {\bibfnamefont {M.}~\bibnamefont
  {{Kramer}}},\ }\href@noop {} {\emph {\bibinfo {title} {Handbook of Pulsar
  Astronomy, by D.~R.~Lorimer , M.~Kramer, Cambridge, UK: Cambridge University
  Press, 2012}}}\ (\bibinfo {year} {2012})\BibitemShut {NoStop}%
\bibitem [{\citenamefont {{Freire}}\ \emph {et~al.}(2012)\citenamefont
  {{Freire}}, \citenamefont {{Wex}}, \citenamefont {{Esposito-Far{\`e}se}},
  \citenamefont {{Verbiest}}, \citenamefont {{Bailes}}, \citenamefont
  {{Jacoby}}, \citenamefont {{Kramer}}, \citenamefont {{Stairs}}, \citenamefont
  {{Antoniadis}},\ and\ \citenamefont {{Janssen}}}]{Freire2012}%
  \BibitemOpen
  \bibfield  {author} {\bibinfo {author} {\bibfnamefont {P.~C.~C.}\
  \bibnamefont {{Freire}}}, \bibinfo {author} {\bibfnamefont {N.}~\bibnamefont
  {{Wex}}}, \bibinfo {author} {\bibfnamefont {G.}~\bibnamefont
  {{Esposito-Far{\`e}se}}}, \bibinfo {author} {\bibfnamefont {J.~P.~W.}\
  \bibnamefont {{Verbiest}}}, \bibinfo {author} {\bibfnamefont
  {M.}~\bibnamefont {{Bailes}}}, \bibinfo {author} {\bibfnamefont {B.~A.}\
  \bibnamefont {{Jacoby}}}, \bibinfo {author} {\bibfnamefont {M.}~\bibnamefont
  {{Kramer}}}, \bibinfo {author} {\bibfnamefont {I.~H.}\ \bibnamefont
  {{Stairs}}}, \bibinfo {author} {\bibfnamefont {J.}~\bibnamefont
  {{Antoniadis}}}, \ and\ \bibinfo {author} {\bibfnamefont {G.~H.}\
  \bibnamefont {{Janssen}}},\ }\href {\doibase
  10.1111/j.1365-2966.2012.21253.x} {\bibfield  {journal} {\bibinfo  {journal}
  {Mon. Not. R. Astron. Soc.}\ }\textbf {\bibinfo {volume} {423}},\ \bibinfo
  {pages} {3328} (\bibinfo {year} {2012})},\ \Eprint
  {http://arxiv.org/abs/1205.1450} {arXiv:1205.1450 [astro-ph.GA]} \BibitemShut
  {NoStop}%
\bibitem [{\citenamefont {{Planck Collaboration}}\ \emph
  {et~al.}(2016)\citenamefont {{Planck Collaboration}}, \citenamefont {{Ade}},
  \citenamefont {{Aghanim}}, \citenamefont {{Arnaud}}, \citenamefont
  {{Ashdown}}, \citenamefont {{Aumont}}, \citenamefont {{Baccigalupi}},
  \citenamefont {{Banday}}, \citenamefont {{Barreiro}}, \citenamefont
  {{Bartlett}},\ and\ \citenamefont {et~al.}}]{Planck2016a}%
  \BibitemOpen
  \bibfield  {author} {\bibinfo {author} {\bibnamefont {{Planck
  Collaboration}}}, \bibinfo {author} {\bibfnamefont {P.~A.~R.}\ \bibnamefont
  {{Ade}}}, \bibinfo {author} {\bibfnamefont {N.}~\bibnamefont {{Aghanim}}},
  \bibinfo {author} {\bibfnamefont {M.}~\bibnamefont {{Arnaud}}}, \bibinfo
  {author} {\bibfnamefont {M.}~\bibnamefont {{Ashdown}}}, \bibinfo {author}
  {\bibfnamefont {J.}~\bibnamefont {{Aumont}}}, \bibinfo {author}
  {\bibfnamefont {C.}~\bibnamefont {{Baccigalupi}}}, \bibinfo {author}
  {\bibfnamefont {A.~J.}\ \bibnamefont {{Banday}}}, \bibinfo {author}
  {\bibfnamefont {R.~B.}\ \bibnamefont {{Barreiro}}}, \bibinfo {author}
  {\bibfnamefont {J.~G.}\ \bibnamefont {{Bartlett}}}, \ and\ \bibinfo {author}
  {\bibnamefont {et~al.}},\ }\href {\doibase 10.1051/0004-6361/201525830}
  {\bibfield  {journal} {\bibinfo  {journal} {Astron. Astrophys.}\ }\textbf
  {\bibinfo {volume} {594}},\ \bibinfo {eid} {A13} (\bibinfo {year} {2016})},\
  \Eprint {http://arxiv.org/abs/1502.01589} {arXiv:1502.01589} \BibitemShut
  {NoStop}%
\bibitem [{\citenamefont {{BICEP2 Collaboration}}\ \emph
  {et~al.}(2014)\citenamefont {{BICEP2 Collaboration}}, \citenamefont {{Ade}},
  \citenamefont {{Aikin}}, \citenamefont {{Barkats}}, \citenamefont {{Benton}},
  \citenamefont {{Bischoff}}, \citenamefont {{Bock}}, \citenamefont {{Brevik}},
  \citenamefont {{Buder}}, \citenamefont {{Bullock}}, \citenamefont {{Dowell}},
  \citenamefont {{Duband}}, \citenamefont {{Filippini}}, \citenamefont
  {{Fliescher}}, \citenamefont {{Golwala}}, \citenamefont {{Halpern}},\ and\
  \citenamefont {et~al.}}]{BICEP2014}%
  \BibitemOpen
  \bibfield  {author} {\bibinfo {author} {\bibnamefont {{BICEP2
  Collaboration}}}, \bibinfo {author} {\bibfnamefont {P.~A.~R.}\ \bibnamefont
  {{Ade}}}, \bibinfo {author} {\bibfnamefont {R.~W.}\ \bibnamefont {{Aikin}}},
  \bibinfo {author} {\bibfnamefont {D.}~\bibnamefont {{Barkats}}}, \bibinfo
  {author} {\bibfnamefont {S.~J.}\ \bibnamefont {{Benton}}}, \bibinfo {author}
  {\bibfnamefont {C.~A.}\ \bibnamefont {{Bischoff}}}, \bibinfo {author}
  {\bibfnamefont {J.~J.}\ \bibnamefont {{Bock}}}, \bibinfo {author}
  {\bibfnamefont {J.~A.}\ \bibnamefont {{Brevik}}}, \bibinfo {author}
  {\bibfnamefont {I.}~\bibnamefont {{Buder}}}, \bibinfo {author} {\bibfnamefont
  {E.}~\bibnamefont {{Bullock}}}, \bibinfo {author} {\bibfnamefont {C.~D.}\
  \bibnamefont {{Dowell}}}, \bibinfo {author} {\bibfnamefont {L.}~\bibnamefont
  {{Duband}}}, \bibinfo {author} {\bibfnamefont {J.~P.}\ \bibnamefont
  {{Filippini}}}, \bibinfo {author} {\bibfnamefont {S.}~\bibnamefont
  {{Fliescher}}}, \bibinfo {author} {\bibfnamefont {S.~R.}\ \bibnamefont
  {{Golwala}}}, \bibinfo {author} {\bibfnamefont {M.}~\bibnamefont
  {{Halpern}}}, \ and\ \bibinfo {author} {\bibnamefont {et~al.}},\ }\href
  {\doibase 10.1103/PhysRevLett.112.241101} {\bibfield  {journal} {\bibinfo
  {journal} {Phys. Rev. Lett.}\ }\textbf {\bibinfo {volume} {112}},\ \bibinfo
  {eid} {241101} (\bibinfo {year} {2014})},\ \Eprint
  {http://arxiv.org/abs/1403.3985} {arXiv:1403.3985} \BibitemShut {NoStop}%
\bibitem [{\citenamefont {{Zwicky}}(1933)}]{Zwicky1933}%
  \BibitemOpen
  \bibfield  {author} {\bibinfo {author} {\bibfnamefont {F.}~\bibnamefont
  {{Zwicky}}},\ }\href@noop {} {\bibfield  {journal} {\bibinfo  {journal}
  {Helvetica Physica Acta}\ }\textbf {\bibinfo {volume} {6}},\ \bibinfo {pages}
  {110} (\bibinfo {year} {1933})}\BibitemShut {NoStop}%
\bibitem [{\citenamefont {{Capozziello}}\ and\ \citenamefont {{De
  Laurentis}}(2012)}]{Capozziello2012}%
  \BibitemOpen
  \bibfield  {author} {\bibinfo {author} {\bibfnamefont {S.}~\bibnamefont
  {{Capozziello}}}\ and\ \bibinfo {author} {\bibfnamefont {M.}~\bibnamefont
  {{De Laurentis}}},\ }\href {\doibase 10.1002/andp.201200109} {\bibfield
  {journal} {\bibinfo  {journal} {Annalen der Physik}\ }\textbf {\bibinfo
  {volume} {524}},\ \bibinfo {pages} {545} (\bibinfo {year}
  {2012})}\BibitemShut {NoStop}%
\bibitem [{\citenamefont {{Johannsen}}\ and\ \citenamefont
  {{Psaltis}}(2011)}]{Johannsen2011}%
  \BibitemOpen
  \bibfield  {author} {\bibinfo {author} {\bibfnamefont {T.}~\bibnamefont
  {{Johannsen}}}\ and\ \bibinfo {author} {\bibfnamefont {D.}~\bibnamefont
  {{Psaltis}}},\ }\href {\doibase 10.1103/PhysRevD.83.124015} {\bibfield
  {journal} {\bibinfo  {journal} {Phys. Rev. D}\ }\textbf {\bibinfo {volume}
  {83}},\ \bibinfo {eid} {124015} (\bibinfo {year} {2011})},\ \Eprint
  {http://arxiv.org/abs/1105.3191} {arXiv:1105.3191 [gr-qc]} \BibitemShut
  {NoStop}%
\bibitem [{\citenamefont {{Lin}}\ \emph {et~al.}(2015)\citenamefont {{Lin}},
  \citenamefont {{Tsukamoto}}, \citenamefont {{Ghasemi-Nodehi}},\ and\
  \citenamefont {{Bambi}}}]{LinBambi2015}%
  \BibitemOpen
  \bibfield  {author} {\bibinfo {author} {\bibfnamefont {N.}~\bibnamefont
  {{Lin}}}, \bibinfo {author} {\bibfnamefont {N.}~\bibnamefont {{Tsukamoto}}},
  \bibinfo {author} {\bibfnamefont {M.}~\bibnamefont {{Ghasemi-Nodehi}}}, \
  and\ \bibinfo {author} {\bibfnamefont {C.}~\bibnamefont {{Bambi}}},\ }\href
  {\doibase 10.1140/epjc/s10052-015-3837-3} {\bibfield  {journal} {\bibinfo
  {journal} {European Physical Journal C}\ }\textbf {\bibinfo {volume} {75}},\
  \bibinfo {eid} {599} (\bibinfo {year} {2015})},\ \Eprint
  {http://arxiv.org/abs/1512.00724} {arXiv:1512.00724 [gr-qc]} \BibitemShut
  {NoStop}%
\bibitem [{\citenamefont {{Rezzolla}}\ and\ \citenamefont
  {{Zhidenko}}(2014)}]{Rezzolla2014}%
  \BibitemOpen
  \bibfield  {author} {\bibinfo {author} {\bibfnamefont {L.}~\bibnamefont
  {{Rezzolla}}}\ and\ \bibinfo {author} {\bibfnamefont {A.}~\bibnamefont
  {{Zhidenko}}},\ }\href {\doibase 10.1103/PhysRevD.90.084009} {\bibfield
  {journal} {\bibinfo  {journal} {Phys. Rev. D}\ }\textbf {\bibinfo {volume}
  {90}},\ \bibinfo {eid} {084009} (\bibinfo {year} {2014})},\ \Eprint
  {http://arxiv.org/abs/1407.3086} {arXiv:1407.3086 [gr-qc]} \BibitemShut
  {NoStop}%
\bibitem [{\citenamefont {{Konoplya}}\ \emph {et~al.}(2016)\citenamefont
  {{Konoplya}}, \citenamefont {{Rezzolla}},\ and\ \citenamefont
  {{Zhidenko}}}]{Konoplya2016a}%
  \BibitemOpen
  \bibfield  {author} {\bibinfo {author} {\bibfnamefont {R.}~\bibnamefont
  {{Konoplya}}}, \bibinfo {author} {\bibfnamefont {L.}~\bibnamefont
  {{Rezzolla}}}, \ and\ \bibinfo {author} {\bibfnamefont {A.}~\bibnamefont
  {{Zhidenko}}},\ }\href {\doibase 10.1103/PhysRevD.93.064015} {\bibfield
  {journal} {\bibinfo  {journal} {Phys. Rev. D}\ }\textbf {\bibinfo {volume}
  {93}},\ \bibinfo {eid} {064015} (\bibinfo {year} {2016})},\ \Eprint
  {http://arxiv.org/abs/1602.02378} {arXiv:1602.02378 [gr-qc]} \BibitemShut
  {NoStop}%
\bibitem [{\citenamefont {{Mizuno}}\ and\ \citenamefont {{\it et
  al.}}(2017)}]{Mizuno2017}%
  \BibitemOpen
  \bibfield  {author} {\bibinfo {author} {\bibfnamefont {Y.}~\bibnamefont
  {{Mizuno}}}\ and\ \bibinfo {author} {\bibnamefont {{\it et al.}}},\
  }\href@noop {} {\  (\bibinfo {year} {2017})},\ \bibinfo {note} {in
  preparation}\BibitemShut {NoStop}%
\bibitem [{\citenamefont {{Kokkotas}}\ \emph {et~al.}(2017)\citenamefont
  {{Kokkotas}}, \citenamefont {{Konoplya}},\ and\ \citenamefont
  {{Zhidenko}}}]{Kokkotas2017}%
  \BibitemOpen
  \bibfield  {author} {\bibinfo {author} {\bibfnamefont {K.~D.}\ \bibnamefont
  {{Kokkotas}}}, \bibinfo {author} {\bibfnamefont {R.~A.}\ \bibnamefont
  {{Konoplya}}}, \ and\ \bibinfo {author} {\bibfnamefont {A.}~\bibnamefont
  {{Zhidenko}}},\ }\href {\doibase 10.1103/PhysRevD.96.064007} {\bibfield
  {journal} {\bibinfo  {journal} {\prd}\ }\textbf {\bibinfo {volume} {96}},\
  \bibinfo {eid} {064007} (\bibinfo {year} {2017})},\ \Eprint
  {http://arxiv.org/abs/1705.09875} {arXiv:1705.09875 [gr-qc]} \BibitemShut
  {NoStop}%
\bibitem [{\citenamefont {Chandrasekhar}(1983)}]{Chandrasekhar83}%
  \BibitemOpen
  \bibfield  {author} {\bibinfo {author} {\bibfnamefont {S.}~\bibnamefont
  {Chandrasekhar}},\ }\href@noop {} {\emph {\bibinfo {title} {The Mathematical
  Theory of Black Holes}}},\ Chandrasekhar83\ (\bibinfo  {publisher} {Oxford
  University Press},\ \bibinfo {address} {Oxford, England},\ \bibinfo {year}
  {1983})\BibitemShut {NoStop}%
\bibitem [{\citenamefont {{Nojiri}}\ and\ \citenamefont
  {{Odintsov}}(2011)}]{Nojiri2011}%
  \BibitemOpen
  \bibfield  {author} {\bibinfo {author} {\bibfnamefont {S.}~\bibnamefont
  {{Nojiri}}}\ and\ \bibinfo {author} {\bibfnamefont {S.~D.}\ \bibnamefont
  {{Odintsov}}},\ }\href {\doibase 10.1140/epjc/s10052-011-1801-4} {\bibfield
  {journal} {\bibinfo  {journal} {European Physical Journal C}\ }\textbf
  {\bibinfo {volume} {71}},\ \bibinfo {eid} {1801} (\bibinfo {year} {2011})},\
  \Eprint {http://arxiv.org/abs/1110.0889} {arXiv:1110.0889 [hep-ph]}
  \BibitemShut {NoStop}%
\bibitem [{\citenamefont {{Cai}}\ \emph {et~al.}(2016)\citenamefont {{Cai}},
  \citenamefont {{Capozziello}}, \citenamefont {{De Laurentis}},\ and\
  \citenamefont {{Saridakis}}}]{Cai2016}%
  \BibitemOpen
  \bibfield  {author} {\bibinfo {author} {\bibfnamefont {Y.-F.}\ \bibnamefont
  {{Cai}}}, \bibinfo {author} {\bibfnamefont {S.}~\bibnamefont
  {{Capozziello}}}, \bibinfo {author} {\bibfnamefont {M.}~\bibnamefont {{De
  Laurentis}}}, \ and\ \bibinfo {author} {\bibfnamefont {E.~N.}\ \bibnamefont
  {{Saridakis}}},\ }\href {\doibase 10.1088/0034-4885/79/10/106901} {\bibfield
  {journal} {\bibinfo  {journal} {Reports on Progress in Physics}\ }\textbf
  {\bibinfo {volume} {79}},\ \bibinfo {eid} {106901} (\bibinfo {year}
  {2016})},\ \Eprint {http://arxiv.org/abs/1511.07586} {arXiv:1511.07586
  [gr-qc]} \BibitemShut {NoStop}%
\bibitem [{\citenamefont {{Brans}}\ and\ \citenamefont
  {{Dicke}}(1961)}]{Brans1961}%
  \BibitemOpen
  \bibfield  {author} {\bibinfo {author} {\bibfnamefont {C.}~\bibnamefont
  {{Brans}}}\ and\ \bibinfo {author} {\bibfnamefont {R.~H.}\ \bibnamefont
  {{Dicke}}},\ }\href {\doibase 10.1103/PhysRev.124.925} {\bibfield  {journal}
  {\bibinfo  {journal} {Physical Review}\ }\textbf {\bibinfo {volume} {124}},\
  \bibinfo {pages} {925} (\bibinfo {year} {1961})}\BibitemShut {NoStop}%
\bibitem [{\citenamefont {{Damour}}(1988)}]{Damour1988}%
  \BibitemOpen
  \bibfield  {author} {\bibinfo {author} {\bibfnamefont {T.}~\bibnamefont
  {{Damour}}},\ }in\ \href@noop {} {\emph {\bibinfo {booktitle} {Proceedings of
  the 2nd Canadian Conference on General Relativity and Relativistic
  Astrophysics}}},\ \bibinfo {editor} {edited by\ \bibinfo {editor}
  {\bibfnamefont {A.}~\bibnamefont {{Coley}}}, \bibinfo {editor} {\bibfnamefont
  {C.}~\bibnamefont {{Dyer}}}, \ and\ \bibinfo {editor} {\bibfnamefont
  {T.}~\bibnamefont {{Tupper}}}}\ (\bibinfo {year} {1988})\ pp.\ \bibinfo
  {pages} {315--334}\BibitemShut {NoStop}%
\bibitem [{\citenamefont {{Damour}}\ and\ \citenamefont
  {{Esposito-Far{\`e}se}}(1998)}]{Damour1998}%
  \BibitemOpen
  \bibfield  {author} {\bibinfo {author} {\bibfnamefont {T.}~\bibnamefont
  {{Damour}}}\ and\ \bibinfo {author} {\bibfnamefont {G.}~\bibnamefont
  {{Esposito-Far{\`e}se}}},\ }\href {\doibase 10.1103/PhysRevD.58.042001}
  {\bibfield  {journal} {\bibinfo  {journal} {Phys. Rev. D}\ }\textbf {\bibinfo
  {volume} {58}},\ \bibinfo {eid} {042001} (\bibinfo {year} {1998})},\ \Eprint
  {http://arxiv.org/abs/gr-qc/9803031} {gr-qc/9803031} \BibitemShut {NoStop}%
\bibitem [{\citenamefont {{Thorne}}\ and\ \citenamefont
  {{Dykla}}(1971)}]{Thorne1971}%
  \BibitemOpen
  \bibfield  {author} {\bibinfo {author} {\bibfnamefont {K.~S.}\ \bibnamefont
  {{Thorne}}}\ and\ \bibinfo {author} {\bibfnamefont {J.~J.}\ \bibnamefont
  {{Dykla}}},\ }\href {\doibase 10.1086/180734} {\bibfield  {journal} {\bibinfo
   {journal} {Astrophys. J. Lett.}\ }\textbf {\bibinfo {volume} {166}},\
  \bibinfo {pages} {L35} (\bibinfo {year} {1971})}\BibitemShut {NoStop}%
\bibitem [{\citenamefont {{Kim}}(1999)}]{Kim1999}%
  \BibitemOpen
  \bibfield  {author} {\bibinfo {author} {\bibfnamefont {H.}~\bibnamefont
  {{Kim}}},\ }\href {\doibase 10.1103/PhysRevD.60.024001} {\bibfield  {journal}
  {\bibinfo  {journal} {Phys. Rev. D}\ }\textbf {\bibinfo {volume} {60}},\
  \bibinfo {eid} {024001} (\bibinfo {year} {1999})},\ \Eprint
  {http://arxiv.org/abs/gr-qc/9811012} {gr-qc/9811012} \BibitemShut {NoStop}%
\bibitem [{\citenamefont {{Capozziello}}\ and\ \citenamefont
  {{Francaviglia}}(2008)}]{Capozziello2008}%
  \BibitemOpen
  \bibfield  {author} {\bibinfo {author} {\bibfnamefont {S.}~\bibnamefont
  {{Capozziello}}}\ and\ \bibinfo {author} {\bibfnamefont {M.}~\bibnamefont
  {{Francaviglia}}},\ }\href {\doibase 10.1007/s10714-007-0551-y} {\bibfield
  {journal} {\bibinfo  {journal} {General Relativity and Gravitation}\ }\textbf
  {\bibinfo {volume} {40}},\ \bibinfo {pages} {357} (\bibinfo {year} {2008})},\
  \Eprint {http://arxiv.org/abs/0706.1146} {arXiv:0706.1146} \BibitemShut
  {NoStop}%
\bibitem [{\citenamefont {{Will}}(1993)}]{Will92}%
  \BibitemOpen
  \bibfield  {author} {\bibinfo {author} {\bibfnamefont {C.~M.}\ \bibnamefont
  {{Will}}},\ }\href {\doibase 10.1017/cbo9780511564246} {\emph {\bibinfo
  {title} {Theory and Experiment in Gravitational Physics, by Clifford M.~Will,
  pp.~396.~ISBN 0521439736.~Cambridge, UK: Cambridge University Press, March
  1993.}}}\ (\bibinfo {year} {1993})\BibitemShut {NoStop}%
\bibitem [{\citenamefont {{Eddington}}(1923)}]{Eddington1923}%
  \BibitemOpen
  \bibfield  {author} {\bibinfo {author} {\bibfnamefont {A.~S.}\ \bibnamefont
  {{Eddington}}},\ }\href@noop {} {\emph {\bibinfo {title} {{The mathematical
  theory of relativity}}}}\ (\bibinfo  {publisher} {Cambridge University
  Press},\ \bibinfo {year} {1923})\BibitemShut {NoStop}%
\bibitem [{\citenamefont {{Capozziello}}\ and\ \citenamefont
  {{Troisi}}(2005)}]{Capozziello2005}%
  \BibitemOpen
  \bibfield  {author} {\bibinfo {author} {\bibfnamefont {S.}~\bibnamefont
  {{Capozziello}}}\ and\ \bibinfo {author} {\bibfnamefont {A.}~\bibnamefont
  {{Troisi}}},\ }\href {\doibase 10.1103/PhysRevD.72.044022} {\bibfield
  {journal} {\bibinfo  {journal} {Phys. Rev. D}\ }\textbf {\bibinfo {volume}
  {72}},\ \bibinfo {eid} {044022} (\bibinfo {year} {2005})},\ \Eprint
  {http://arxiv.org/abs/astro-ph/0507545} {astro-ph/0507545} \BibitemShut
  {NoStop}%
\bibitem [{\citenamefont {{Capozziello}}\ \emph {et~al.}(2006)\citenamefont
  {{Capozziello}}, \citenamefont {{Stabile}},\ and\ \citenamefont
  {{Troisi}}}]{Capozziello2006}%
  \BibitemOpen
  \bibfield  {author} {\bibinfo {author} {\bibfnamefont {S.}~\bibnamefont
  {{Capozziello}}}, \bibinfo {author} {\bibfnamefont {A.}~\bibnamefont
  {{Stabile}}}, \ and\ \bibinfo {author} {\bibfnamefont {A.}~\bibnamefont
  {{Troisi}}},\ }\href {\doibase 10.1142/S0217732306021633} {\bibfield
  {journal} {\bibinfo  {journal} {Modern Physics Letters A}\ }\textbf {\bibinfo
  {volume} {21}},\ \bibinfo {pages} {2291} (\bibinfo {year} {2006})},\ \Eprint
  {http://arxiv.org/abs/gr-qc/0603071} {gr-qc/0603071} \BibitemShut {NoStop}%
\bibitem [{\citenamefont {{Capozziello}}\ \emph {et~al.}(2009)\citenamefont
  {{Capozziello}}, \citenamefont {{de Laurentis}}, \citenamefont {{Nojiri}},\
  and\ \citenamefont {{Odintsov}}}]{Capozziello2009}%
  \BibitemOpen
  \bibfield  {author} {\bibinfo {author} {\bibfnamefont {S.}~\bibnamefont
  {{Capozziello}}}, \bibinfo {author} {\bibfnamefont {M.}~\bibnamefont {{de
  Laurentis}}}, \bibinfo {author} {\bibfnamefont {S.}~\bibnamefont {{Nojiri}}},
  \ and\ \bibinfo {author} {\bibfnamefont {S.~D.}\ \bibnamefont {{Odintsov}}},\
  }\href {\doibase 10.1007/s10714-009-0758-1} {\bibfield  {journal} {\bibinfo
  {journal} {General Relativity and Gravitation}\ }\textbf {\bibinfo {volume}
  {41}},\ \bibinfo {pages} {2313} (\bibinfo {year} {2009})},\ \Eprint
  {http://arxiv.org/abs/0808.1335} {arXiv:0808.1335 [hep-th]} \BibitemShut
  {NoStop}%
\bibitem [{\citenamefont {{Damour}}(2007)}]{Damour2007}%
  \BibitemOpen
  \bibfield  {author} {\bibinfo {author} {\bibfnamefont {T.}~\bibnamefont
  {{Damour}}},\ }\href@noop {} {\bibfield  {journal} {\bibinfo  {journal}
  {ArXiv e-prints}\ } (\bibinfo {year} {2007})},\ \Eprint
  {http://arxiv.org/abs/0705.3109} {arXiv:0705.3109 [gr-qc]} \BibitemShut
  {NoStop}%
\bibitem [{\citenamefont {{De Laurentis}}\ \emph {et~al.}(2015)\citenamefont
  {{De Laurentis}}, \citenamefont {{Paolella}},\ and\ \citenamefont
  {{Capozziello}}}]{DeLaurentis2015}%
  \BibitemOpen
  \bibfield  {author} {\bibinfo {author} {\bibfnamefont {M.}~\bibnamefont {{De
  Laurentis}}}, \bibinfo {author} {\bibfnamefont {M.}~\bibnamefont
  {{Paolella}}}, \ and\ \bibinfo {author} {\bibfnamefont {S.}~\bibnamefont
  {{Capozziello}}},\ }\href {\doibase 10.1103/PhysRevD.91.083531} {\bibfield
  {journal} {\bibinfo  {journal} {Phys. Rev. D}\ }\textbf {\bibinfo {volume}
  {91}},\ \bibinfo {eid} {083531} (\bibinfo {year} {2015})},\ \Eprint
  {http://arxiv.org/abs/1503.04659} {arXiv:1503.04659 [gr-qc]} \BibitemShut
  {NoStop}%
\bibitem [{\citenamefont {Birell}\ and\ \citenamefont
  {Davis}(1982)}]{Birell82}%
  \BibitemOpen
  \bibfield  {author} {\bibinfo {author} {\bibfnamefont {N.~D.}\ \bibnamefont
  {Birell}}\ and\ \bibinfo {author} {\bibfnamefont {P.~C.~W.}\ \bibnamefont
  {Davis}},\ }\href@noop {} {\emph {\bibinfo {title} {Quantum fields in curved
  space}}}\ (\bibinfo  {publisher} {Cambridge University Press},\ \bibinfo
  {address} {Cambridge},\ \bibinfo {year} {1982})\BibitemShut {NoStop}%
\bibitem [{\citenamefont {{Borka}}\ \emph {et~al.}(2016)\citenamefont
  {{Borka}}, \citenamefont {{Capozziello}}, \citenamefont {{Jovanovi{\'c}}},\
  and\ \citenamefont {{Borka Jovanovi{\'c}}}}]{Borka2016}%
  \BibitemOpen
  \bibfield  {author} {\bibinfo {author} {\bibfnamefont {D.}~\bibnamefont
  {{Borka}}}, \bibinfo {author} {\bibfnamefont {S.}~\bibnamefont
  {{Capozziello}}}, \bibinfo {author} {\bibfnamefont {P.}~\bibnamefont
  {{Jovanovi{\'c}}}}, \ and\ \bibinfo {author} {\bibfnamefont {V.}~\bibnamefont
  {{Borka Jovanovi{\'c}}}},\ }\href {\doibase
  10.1016/j.astropartphys.2016.03.002} {\bibfield  {journal} {\bibinfo
  {journal} {Astroparticle Physics}\ }\textbf {\bibinfo {volume} {79}},\
  \bibinfo {pages} {41} (\bibinfo {year} {2016})},\ \Eprint
  {http://arxiv.org/abs/1504.07832} {arXiv:1504.07832 [gr-qc]} \BibitemShut
  {NoStop}%
\bibitem [{\citenamefont {{Capozziello}}\ \emph {et~al.}(2014)\citenamefont
  {{Capozziello}}, \citenamefont {{Borka}}, \citenamefont {{Jovanovi{\'c}}},\
  and\ \citenamefont {{Jovanovi{\'c}}}}]{Capozziello2014}%
  \BibitemOpen
  \bibfield  {author} {\bibinfo {author} {\bibfnamefont {S.}~\bibnamefont
  {{Capozziello}}}, \bibinfo {author} {\bibfnamefont {D.}~\bibnamefont
  {{Borka}}}, \bibinfo {author} {\bibfnamefont {P.}~\bibnamefont
  {{Jovanovi{\'c}}}}, \ and\ \bibinfo {author} {\bibfnamefont {V.~B.}\
  \bibnamefont {{Jovanovi{\'c}}}},\ }\href {\doibase
  10.1103/PhysRevD.90.044052} {\bibfield  {journal} {\bibinfo  {journal} {Phys.
  Rev. D}\ }\textbf {\bibinfo {volume} {90}},\ \bibinfo {eid} {044052}
  (\bibinfo {year} {2014})},\ \Eprint {http://arxiv.org/abs/1408.1169}
  {arXiv:1408.1169} \BibitemShut {NoStop}%
\bibitem [{\citenamefont {{Hees}}\ \emph {et~al.}(2017)\citenamefont {{Hees}},
  \citenamefont {{Ghez}}, \citenamefont {{Do}}, \citenamefont {{Lu}},
  \citenamefont {{Morris}}, \citenamefont {{Becklin}}, \citenamefont
  {{Witzel}}, \citenamefont {{Boehle}}, \citenamefont {{Chappell}},
  \citenamefont {{Chen}}, \citenamefont {{Chu}}, \citenamefont {{Ciurlo}},
  \citenamefont {{Dehghanfar}}, \citenamefont {{Gallego-Cano}}, \citenamefont
  {{Gautam}},\ and\ \citenamefont {et~al.}}]{Hees2017}%
  \BibitemOpen
  \bibfield  {author} {\bibinfo {author} {\bibfnamefont {A.}~\bibnamefont
  {{Hees}}}, \bibinfo {author} {\bibfnamefont {A.~M.}\ \bibnamefont {{Ghez}}},
  \bibinfo {author} {\bibfnamefont {T.}~\bibnamefont {{Do}}}, \bibinfo {author}
  {\bibfnamefont {J.~R.}\ \bibnamefont {{Lu}}}, \bibinfo {author}
  {\bibfnamefont {M.~R.}\ \bibnamefont {{Morris}}}, \bibinfo {author}
  {\bibfnamefont {E.~E.}\ \bibnamefont {{Becklin}}}, \bibinfo {author}
  {\bibfnamefont {G.}~\bibnamefont {{Witzel}}}, \bibinfo {author}
  {\bibfnamefont {A.}~\bibnamefont {{Boehle}}}, \bibinfo {author}
  {\bibfnamefont {S.}~\bibnamefont {{Chappell}}}, \bibinfo {author}
  {\bibfnamefont {Z.}~\bibnamefont {{Chen}}}, \bibinfo {author} {\bibfnamefont
  {D.}~\bibnamefont {{Chu}}}, \bibinfo {author} {\bibfnamefont
  {A.}~\bibnamefont {{Ciurlo}}}, \bibinfo {author} {\bibfnamefont
  {A.}~\bibnamefont {{Dehghanfar}}}, \bibinfo {author} {\bibfnamefont
  {E.}~\bibnamefont {{Gallego-Cano}}}, \bibinfo {author} {\bibfnamefont
  {A.}~\bibnamefont {{Gautam}}}, \ and\ \bibinfo {author} {\bibnamefont
  {et~al.}},\ }\href@noop {} {\bibfield  {journal} {\bibinfo  {journal} {ArXiv
  e-prints}\ } (\bibinfo {year} {2017})},\ \Eprint
  {http://arxiv.org/abs/1705.10792} {arXiv:1705.10792} \BibitemShut {NoStop}%
\bibitem [{\citenamefont {{Garc{\'{\i}}a}}\ \emph {et~al.}(1995)\citenamefont
  {{Garc{\'{\i}}a}}, \citenamefont {{Galtsov}},\ and\ \citenamefont
  {{Kechkin}}}]{Garcia1995}%
  \BibitemOpen
  \bibfield  {author} {\bibinfo {author} {\bibfnamefont {A.}~\bibnamefont
  {{Garc{\'{\i}}a}}}, \bibinfo {author} {\bibfnamefont {D.}~\bibnamefont
  {{Galtsov}}}, \ and\ \bibinfo {author} {\bibfnamefont {O.}~\bibnamefont
  {{Kechkin}}},\ }\href {\doibase 10.1103/PhysRevLett.74.1276} {\bibfield
  {journal} {\bibinfo  {journal} {Phys. Rev. Lett.}\ }\textbf {\bibinfo
  {volume} {74}},\ \bibinfo {pages} {1276} (\bibinfo {year}
  {1995})}\BibitemShut {NoStop}%
\bibitem [{\citenamefont {{Gibbons}}\ and\ \citenamefont
  {{Maeda}}(1988)}]{Gibbons1988}%
  \BibitemOpen
  \bibfield  {author} {\bibinfo {author} {\bibfnamefont {G.~W.}\ \bibnamefont
  {{Gibbons}}}\ and\ \bibinfo {author} {\bibfnamefont {K.-I.}\ \bibnamefont
  {{Maeda}}},\ }\href {\doibase 10.1016/0550-3213(88)90006-5} {\bibfield
  {journal} {\bibinfo  {journal} {Nuclear Physics B}\ }\textbf {\bibinfo
  {volume} {298}},\ \bibinfo {pages} {741} (\bibinfo {year}
  {1988})}\BibitemShut {NoStop}%
\bibitem [{\citenamefont {{Garfinkle}}\ \emph {et~al.}(1991)\citenamefont
  {{Garfinkle}}, \citenamefont {{Horowitz}},\ and\ \citenamefont
  {{Strominger}}}]{Garfinkle1991}%
  \BibitemOpen
  \bibfield  {author} {\bibinfo {author} {\bibfnamefont {D.}~\bibnamefont
  {{Garfinkle}}}, \bibinfo {author} {\bibfnamefont {G.~T.}\ \bibnamefont
  {{Horowitz}}}, \ and\ \bibinfo {author} {\bibfnamefont {A.}~\bibnamefont
  {{Strominger}}},\ }\href {\doibase 10.1103/PhysRevD.43.3140} {\bibfield
  {journal} {\bibinfo  {journal} {Phys. Rev. D}\ }\textbf {\bibinfo {volume}
  {43}},\ \bibinfo {pages} {3140} (\bibinfo {year} {1991})}\BibitemShut
  {NoStop}%
\bibitem [{\citenamefont {{Horowitz}}\ and\ \citenamefont
  {{Strominger}}(1991)}]{Horowitz1991}%
  \BibitemOpen
  \bibfield  {author} {\bibinfo {author} {\bibfnamefont {G.~T.}\ \bibnamefont
  {{Horowitz}}}\ and\ \bibinfo {author} {\bibfnamefont {A.}~\bibnamefont
  {{Strominger}}},\ }\href {\doibase 10.1016/0550-3213(91)90440-9} {\bibfield
  {journal} {\bibinfo  {journal} {Nuclear Physics B}\ }\textbf {\bibinfo
  {volume} {360}},\ \bibinfo {pages} {197} (\bibinfo {year}
  {1991})}\BibitemShut {NoStop}%
\bibitem [{\citenamefont {{Shapere}}\ \emph {et~al.}(1991)\citenamefont
  {{Shapere}}, \citenamefont {{Trivedi}},\ and\ \citenamefont
  {{Wilczek}}}]{Shapere1991}%
  \BibitemOpen
  \bibfield  {author} {\bibinfo {author} {\bibfnamefont {A.}~\bibnamefont
  {{Shapere}}}, \bibinfo {author} {\bibfnamefont {S.}~\bibnamefont
  {{Trivedi}}}, \ and\ \bibinfo {author} {\bibfnamefont {F.}~\bibnamefont
  {{Wilczek}}},\ }\href {\doibase 10.1142/S0217732391003122} {\bibfield
  {journal} {\bibinfo  {journal} {Modern Physics Letters A}\ }\textbf {\bibinfo
  {volume} {6}},\ \bibinfo {pages} {2677} (\bibinfo {year} {1991})}\BibitemShut
  {NoStop}%
\bibitem [{\citenamefont {{Sen}}(1992)}]{Sen1992}%
  \BibitemOpen
  \bibfield  {author} {\bibinfo {author} {\bibfnamefont {A.}~\bibnamefont
  {{Sen}}},\ }\href {\doibase 10.1103/PhysRevLett.69.1006} {\bibfield
  {journal} {\bibinfo  {journal} {Phys. Rev. Lett.}\ }\textbf {\bibinfo
  {volume} {69}},\ \bibinfo {pages} {1006} (\bibinfo {year} {1992})},\ \Eprint
  {http://arxiv.org/abs/hep-th/9204046} {hep-th/9204046} \BibitemShut {NoStop}%
\bibitem [{\citenamefont {Landau}\ and\ \citenamefont
  {Lifshitz}(1976)}]{Landau-Lifshitz1}%
  \BibitemOpen
  \bibfield  {author} {\bibinfo {author} {\bibfnamefont {L.~D.}\ \bibnamefont
  {Landau}}\ and\ \bibinfo {author} {\bibfnamefont {E.~M.}\ \bibnamefont
  {Lifshitz}},\ }\href@noop {} {\emph {\bibinfo {title} {Mechanics, Volume
  1}}}\ (\bibinfo  {publisher} {Pergamon Press},\ \bibinfo {address} {Oxford},\
  \bibinfo {year} {1976})\BibitemShut {NoStop}%
\bibitem [{\citenamefont {{Roy}}(2005)}]{Roy2005}%
  \BibitemOpen
  \bibfield  {author} {\bibinfo {author} {\bibfnamefont {A.~E.}\ \bibnamefont
  {{Roy}}},\ }\href@noop {} {\emph {\bibinfo {title} {Orbital motion /
  A.~E.~Roy.~Bristol (UK): Institute of Physics Publishing, 4th edition.~ISBN
  0-7503-1015-6, 2005, XVIII + 526 pp.}}}\ (\bibinfo  {publisher} {tute of
  Physics Publishing},\ \bibinfo {year} {2005})\BibitemShut {NoStop}%
\bibitem [{\citenamefont {{Darwin}}(1961)}]{Darwin1961}%
  \BibitemOpen
  \bibfield  {author} {\bibinfo {author} {\bibfnamefont {C.}~\bibnamefont
  {{Darwin}}},\ }\href {\doibase 10.1098/rspa.1961.0142} {\bibfield  {journal}
  {\bibinfo  {journal} {Proceedings of the Royal Society of London Series A}\
  }\textbf {\bibinfo {volume} {263}},\ \bibinfo {pages} {39} (\bibinfo {year}
  {1961})}\BibitemShut {NoStop}%
\bibitem [{\citenamefont {{Geisler}}(1963)}]{Geisler1963}%
  \BibitemOpen
  \bibfield  {author} {\bibinfo {author} {\bibfnamefont {P.~A.}\ \bibnamefont
  {{Geisler}}},\ }\href {\doibase 10.1086/109199} {\bibfield  {journal}
  {\bibinfo  {journal} {Astron. J.}\ }\textbf {\bibinfo {volume} {68}},\
  \bibinfo {pages} {715} (\bibinfo {year} {1963})}\BibitemShut {NoStop}%
\bibitem [{\citenamefont {Frolov}\ and\ \citenamefont
  {Novikov}(1998)}]{Frolov98}%
  \BibitemOpen
  \bibfield  {author} {\bibinfo {author} {\bibfnamefont {V.~P.}\ \bibnamefont
  {Frolov}}\ and\ \bibinfo {author} {\bibfnamefont {I.~D.}\ \bibnamefont
  {Novikov}},\ }\href@noop {} {\emph {\bibinfo {title} {Black Hole Physics,
  Basic Concepts and New Developments}}},\ Frolov98\ (\bibinfo  {publisher}
  {Kluwer Academic Publishers},\ \bibinfo {address} {Dordrecth, The
  Netherlands},\ \bibinfo {year} {1998})\BibitemShut {NoStop}%
\bibitem [{\citenamefont {Maggiore}(2007)}]{Maggiore2007}%
  \BibitemOpen
  \bibfield  {author} {\bibinfo {author} {\bibfnamefont {M.}~\bibnamefont
  {Maggiore}},\ }\href {http://books.google.de/books?id=AqVpQgAACAAJ} {\emph
  {\bibinfo {title} {Gravitational Waves: Volume 1: Theory and Experiments}}},\
  Gravitational Waves\ (\bibinfo  {publisher} {Oxford University Press, USA},\
  \bibinfo {year} {2007})\BibitemShut {NoStop}%
\bibitem [{\citenamefont {{Damour}}\ and\ \citenamefont
  {{Deruelle}}(1985)}]{Damour1985}%
  \BibitemOpen
  \bibfield  {author} {\bibinfo {author} {\bibfnamefont {T.}~\bibnamefont
  {{Damour}}}\ and\ \bibinfo {author} {\bibfnamefont {N.}~\bibnamefont
  {{Deruelle}}},\ }\href@noop {} {\bibfield  {journal} {\bibinfo  {journal}
  {Ann.~Inst.~Henri Poincar{\'e} Phys.~Th{\'e}or., Vol.~43, No.~1, p.~107 -
  132}\ }\textbf {\bibinfo {volume} {43}},\ \bibinfo {pages} {107} (\bibinfo
  {year} {1985})}\BibitemShut {NoStop}%
\bibitem [{\citenamefont {{Damour}}\ and\ \citenamefont
  {{Deruelle}}(1986)}]{Damour1986}%
  \BibitemOpen
  \bibfield  {author} {\bibinfo {author} {\bibfnamefont {T.}~\bibnamefont
  {{Damour}}}\ and\ \bibinfo {author} {\bibfnamefont {N.}~\bibnamefont
  {{Deruelle}}},\ }\href@noop {} {\bibfield  {journal} {\bibinfo  {journal}
  {Ann.~Inst.~Henri Poincar{\'e} Phys.~Th{\'e}or., Vol.~44, No.~3, p.~263 -
  292}\ }\textbf {\bibinfo {volume} {44}},\ \bibinfo {pages} {263} (\bibinfo
  {year} {1986})}\BibitemShut {NoStop}%
\bibitem [{\citenamefont {{Eckart}}\ and\ \citenamefont
  {{Genzel}}(1997)}]{Eckart1997}%
  \BibitemOpen
  \bibfield  {author} {\bibinfo {author} {\bibfnamefont {A.}~\bibnamefont
  {{Eckart}}}\ and\ \bibinfo {author} {\bibfnamefont {R.}~\bibnamefont
  {{Genzel}}},\ }\href {\doibase 10.1093/mnras/284.3.576} {\bibfield  {journal}
  {\bibinfo  {journal} {Mon. Not. R. Astron. Soc.}\ }\textbf {\bibinfo {volume}
  {284}},\ \bibinfo {pages} {576} (\bibinfo {year} {1997})}\BibitemShut
  {NoStop}%
\bibitem [{\citenamefont {{Gillessen}}\ \emph
  {et~al.}(2009{\natexlab{a}})\citenamefont {{Gillessen}}, \citenamefont
  {{Eisenhauer}}, \citenamefont {{Trippe}}, \citenamefont {{Alexander}},
  \citenamefont {{Genzel}}, \citenamefont {{Martins}},\ and\ \citenamefont
  {{Ott}}}]{Gillessen:2009}%
  \BibitemOpen
  \bibfield  {author} {\bibinfo {author} {\bibfnamefont {S.}~\bibnamefont
  {{Gillessen}}}, \bibinfo {author} {\bibfnamefont {F.}~\bibnamefont
  {{Eisenhauer}}}, \bibinfo {author} {\bibfnamefont {S.}~\bibnamefont
  {{Trippe}}}, \bibinfo {author} {\bibfnamefont {T.}~\bibnamefont
  {{Alexander}}}, \bibinfo {author} {\bibfnamefont {R.}~\bibnamefont
  {{Genzel}}}, \bibinfo {author} {\bibfnamefont {F.}~\bibnamefont {{Martins}}},
  \ and\ \bibinfo {author} {\bibfnamefont {T.}~\bibnamefont {{Ott}}},\ }\href
  {\doibase 10.1088/0004-637X/692/2/1075} {\bibfield  {journal} {\bibinfo
  {journal} {Astrophys. J.}\ }\textbf {\bibinfo {volume} {692}},\ \bibinfo
  {pages} {1075} (\bibinfo {year} {2009}{\natexlab{a}})},\ \Eprint
  {http://arxiv.org/abs/0810.4674} {arXiv:0810.4674} \BibitemShut {NoStop}%
\bibitem [{\citenamefont {{Gillessen}}\ \emph
  {et~al.}(2009{\natexlab{b}})\citenamefont {{Gillessen}}, \citenamefont
  {{Eisenhauer}}, \citenamefont {{Fritz}}, \citenamefont {{Bartko}},
  \citenamefont {{Dodds-Eden}}, \citenamefont {{Pfuhl}}, \citenamefont
  {{Ott}},\ and\ \citenamefont {{Genzel}}}]{Gillessen2009L}%
  \BibitemOpen
  \bibfield  {author} {\bibinfo {author} {\bibfnamefont {S.}~\bibnamefont
  {{Gillessen}}}, \bibinfo {author} {\bibfnamefont {F.}~\bibnamefont
  {{Eisenhauer}}}, \bibinfo {author} {\bibfnamefont {T.~K.}\ \bibnamefont
  {{Fritz}}}, \bibinfo {author} {\bibfnamefont {H.}~\bibnamefont {{Bartko}}},
  \bibinfo {author} {\bibfnamefont {K.}~\bibnamefont {{Dodds-Eden}}}, \bibinfo
  {author} {\bibfnamefont {O.}~\bibnamefont {{Pfuhl}}}, \bibinfo {author}
  {\bibfnamefont {T.}~\bibnamefont {{Ott}}}, \ and\ \bibinfo {author}
  {\bibfnamefont {R.}~\bibnamefont {{Genzel}}},\ }\href {\doibase
  10.1088/0004-637X/707/2/L114} {\bibfield  {journal} {\bibinfo  {journal}
  {Astrophys. J. Lett.}\ }\textbf {\bibinfo {volume} {707}},\ \bibinfo {pages}
  {L114} (\bibinfo {year} {2009}{\natexlab{b}})},\ \Eprint
  {http://arxiv.org/abs/0910.3069} {arXiv:0910.3069 [astro-ph.GA]} \BibitemShut
  {NoStop}%
\bibitem [{\citenamefont {{Ghez}}\ \emph {et~al.}(1998)\citenamefont {{Ghez}},
  \citenamefont {{Klein}}, \citenamefont {{Morris}},\ and\ \citenamefont
  {{Becklin}}}]{Ghez1998}%
  \BibitemOpen
  \bibfield  {author} {\bibinfo {author} {\bibfnamefont {A.~M.}\ \bibnamefont
  {{Ghez}}}, \bibinfo {author} {\bibfnamefont {B.~L.}\ \bibnamefont {{Klein}}},
  \bibinfo {author} {\bibfnamefont {M.}~\bibnamefont {{Morris}}}, \ and\
  \bibinfo {author} {\bibfnamefont {E.~E.}\ \bibnamefont {{Becklin}}},\ }\href
  {\doibase 10.1086/306528} {\bibfield  {journal} {\bibinfo  {journal}
  {Astrophys. J.}\ }\textbf {\bibinfo {volume} {509}},\ \bibinfo {pages} {678}
  (\bibinfo {year} {1998})},\ \Eprint {http://arxiv.org/abs/astro-ph/9807210}
  {astro-ph/9807210} \BibitemShut {NoStop}%
\bibitem [{\citenamefont {{Boehle}}\ \emph {et~al.}(2016)\citenamefont
  {{Boehle}}, \citenamefont {{Ghez}}, \citenamefont {{Sch{\"o}del}},
  \citenamefont {{Meyer}}, \citenamefont {{Yelda}}, \citenamefont {{Albers}},
  \citenamefont {{Martinez}}, \citenamefont {{Becklin}}, \citenamefont {{Do}},
  \citenamefont {{Lu}}, \citenamefont {{Matthews}}, \citenamefont {{Morris}},
  \citenamefont {{Sitarski}},\ and\ \citenamefont {{Witzel}}}]{Boehle2016}%
  \BibitemOpen
  \bibfield  {author} {\bibinfo {author} {\bibfnamefont {A.}~\bibnamefont
  {{Boehle}}}, \bibinfo {author} {\bibfnamefont {A.~M.}\ \bibnamefont
  {{Ghez}}}, \bibinfo {author} {\bibfnamefont {R.}~\bibnamefont
  {{Sch{\"o}del}}}, \bibinfo {author} {\bibfnamefont {L.}~\bibnamefont
  {{Meyer}}}, \bibinfo {author} {\bibfnamefont {S.}~\bibnamefont {{Yelda}}},
  \bibinfo {author} {\bibfnamefont {S.}~\bibnamefont {{Albers}}}, \bibinfo
  {author} {\bibfnamefont {G.~D.}\ \bibnamefont {{Martinez}}}, \bibinfo
  {author} {\bibfnamefont {E.~E.}\ \bibnamefont {{Becklin}}}, \bibinfo {author}
  {\bibfnamefont {T.}~\bibnamefont {{Do}}}, \bibinfo {author} {\bibfnamefont
  {J.~R.}\ \bibnamefont {{Lu}}}, \bibinfo {author} {\bibfnamefont
  {K.}~\bibnamefont {{Matthews}}}, \bibinfo {author} {\bibfnamefont {M.~R.}\
  \bibnamefont {{Morris}}}, \bibinfo {author} {\bibfnamefont {B.}~\bibnamefont
  {{Sitarski}}}, \ and\ \bibinfo {author} {\bibfnamefont {G.}~\bibnamefont
  {{Witzel}}},\ }\href {\doibase 10.3847/0004-637X/830/1/17} {\bibfield
  {journal} {\bibinfo  {journal} {Astrophys. J.}\ }\textbf {\bibinfo {volume}
  {830}},\ \bibinfo {eid} {17} (\bibinfo {year} {2016})},\ \Eprint
  {http://arxiv.org/abs/1607.05726} {arXiv:1607.05726} \BibitemShut {NoStop}%
\bibitem [{\citenamefont {{Liu}}\ \emph {et~al.}(2012)\citenamefont {{Liu}},
  \citenamefont {{Wex}}, \citenamefont {{Kramer}}, \citenamefont {{Cordes}},\
  and\ \citenamefont {{Lazio}}}]{Liu2012}%
  \BibitemOpen
  \bibfield  {author} {\bibinfo {author} {\bibfnamefont {K.}~\bibnamefont
  {{Liu}}}, \bibinfo {author} {\bibfnamefont {N.}~\bibnamefont {{Wex}}},
  \bibinfo {author} {\bibfnamefont {M.}~\bibnamefont {{Kramer}}}, \bibinfo
  {author} {\bibfnamefont {J.~M.}\ \bibnamefont {{Cordes}}}, \ and\ \bibinfo
  {author} {\bibfnamefont {T.~J.~W.}\ \bibnamefont {{Lazio}}},\ }\href
  {\doibase 10.1088/0004-637X/747/1/1} {\bibfield  {journal} {\bibinfo
  {journal} {Astrophys. J.}\ }\textbf {\bibinfo {volume} {747}},\ \bibinfo
  {eid} {1} (\bibinfo {year} {2012})},\ \Eprint
  {http://arxiv.org/abs/1112.2151} {arXiv:1112.2151 [astro-ph.HE]} \BibitemShut
  {NoStop}%
\bibitem [{\citenamefont {{Manchester}}(2013)}]{Manchester2013}%
  \BibitemOpen
  \bibfield  {author} {\bibinfo {author} {\bibfnamefont {R.~N.}\ \bibnamefont
  {{Manchester}}},\ }\href {\doibase 10.1142/S0218271813410071} {\bibfield
  {journal} {\bibinfo  {journal} {International Journal of Modern Physics D}\
  }\textbf {\bibinfo {volume} {22}},\ \bibinfo {eid} {1341007} (\bibinfo {year}
  {2013})}\BibitemShut {NoStop}%
\bibitem [{\citenamefont {{Psaltis}}\ \emph {et~al.}(2016)\citenamefont
  {{Psaltis}}, \citenamefont {{Wex}},\ and\ \citenamefont
  {{Kramer}}}]{Psaltis2016}%
  \BibitemOpen
  \bibfield  {author} {\bibinfo {author} {\bibfnamefont {D.}~\bibnamefont
  {{Psaltis}}}, \bibinfo {author} {\bibfnamefont {N.}~\bibnamefont {{Wex}}}, \
  and\ \bibinfo {author} {\bibfnamefont {M.}~\bibnamefont {{Kramer}}},\ }\href
  {\doibase 10.3847/0004-637X/818/2/121} {\bibfield  {journal} {\bibinfo
  {journal} {Astrophys. J.}\ }\textbf {\bibinfo {volume} {818}},\ \bibinfo
  {eid} {121} (\bibinfo {year} {2016})},\ \Eprint
  {http://arxiv.org/abs/1510.00394} {arXiv:1510.00394 [astro-ph.HE]}
  \BibitemShut {NoStop}%
\end{thebibliography}%

\end{document}